\RequirePackage{lineno}

\documentclass[onecolumn,preprintnumbers,amsmath,amssymb]{revtex4}
\usepackage{graphicx}
\usepackage{bm}
\usepackage{epsfig}
\usepackage{subfigure}
\usepackage{xcolor}

\newcommand{\bi}{\begin{itemize}}
\newcommand{\ei}{\end{itemize}}
\newcommand{\be}{\begin{eqnarray}}
\newcommand{\ee}{\end{eqnarray}}
\newcommand{\beq}{\begin{equation}}
\newcommand{\eeq}{\end{equation}}
\newcommand{\beqn}{\begin{equation*}}
\newcommand{\eeqn}{\end{equation*}}
\newcommand{\bbmatrix}{\left( \begin{array}}
\newcommand{\eematrix}{\end{array} \right)}

\def\blue#1{\textcolor{blue}{#1}}

\def\change#1{\textcolor{black}{#1}}
\def\dd{\text{d}}

\begin{document}

\title{Time-optimal single-scalar control on a  qubit of unitary dynamics  }

\author{Chungwei Lin$^{1}$, Petros T. Boufounos$^{1}$, Yanting Ma$^{1}$, Yebin Wang$^{1}$}

\affiliation{Mitsubishi Electric Research Laboratories (MERL), 201 Broadway, Cambridge, MA 02139}

\author{Qi Ding$^{2}$}
\affiliation{$^2$Department of Electrical Engineering and Computer Science, Massachusetts Institute of Technology, Cambridge, MA 02139, USA}

\author{Dries Sels$^{3,4}$}
\affiliation{
$^3$Department of physics, New York University, New York City, NY 10003, USA \\
$^4$Center for Computational Quantum Physics, Flatiron Institute, 162 5th Ave, New York, NY 10010, USA\footnote{The Flatiron Institute is a division of the Simons Foundation.}
}

\author{Chih-Chun Chien$^{5}$}
\affiliation{Department of physics, University of California, Merced, CA 95343, USA}

\date{\today}
\begin{abstract}
Optimal control theory is applied to analyze the time-optimal solution with a single scalar control knob in a two-level quantum system without quantum decoherence. Emphasis is \change{placed} on the dependence on the maximum control strength $u_\text{max}$.
General constraints on the optimal protocol are derived and used to rigorously parameterize the time-optimal solution. Two concrete problems are investigated.
For generic state preparation problems, both multiple bang-bang and bang-singular-bang are legitimate and should be considered. Generally, the optimal is bang-bang for small $u_\text{max}$, and there exists a state-dependent critical amplitude above which singular control  emerges.
For the X-gate operation of a qubit, the optimal protocol \change{is exclusively} multiple bang-bang. The minimum gate time is about 80\% of that based on the resonant Rabi $\pi$-pulse over a wide range of control strength; in the $u_\text{max} \rightarrow 0$ limit this ratio is derived to be $\pi/4$.
To develop practically feasible protocols, we present methods to smooth the abrupt changes in the bang-bang control while preserving perfect gate fidelity.
\change{The presence of bang-bang segments in the time-optimal protocol} indicates that the high-frequency components and a full calculation (instead of the commonly adopted Rotating Wave Approximation) are essential for the ultimate quantum speed limit. 

\end{abstract}

\maketitle

\section{Introduction}

Optimal Control Theory (OCT), also known as Pontryagin's Maximum Principle (PMP), is a powerful tool to analyze and construct the open-loop optimal control protocol \cite{book:Luenberger, book:Liberzon, book:GeometricOptimalControl, book:Pontryagin}.
The basic formalism of OCT is the calculus of variations, but its general applicability requires detailed analysis that takes the control constraint and the non-smooth behavior into account \cite{book:Liberzon, book:GeometricOptimalControl}.
OCT aims to minimize a user-defined terminal cost function\change{, subject to} the dynamics that contains a time-dependent control protocol, and its success hinges on a sufficiently accurate model due to its open-loop nature.
This framework \change{is naturally suited to} a wide class of quantum tasks \cite{doi:10.1116/5.0006785, PRXQuantum.2.030203, PRXQuantum.2.010101, ansel2024introduction}. The state variables, \change{represented by} the wave function of the quantum system, cannot be completely determined during the evolution while the governing dynamics, the Schr\"odinger equation describing the quantum system, are usually known to high precision.
Well-known examples include the fast quantum state preparation \cite{PhysRevA.85.012317, PhysRevA.97.062343, PhysRevX.8.021012, Pechen_2017, PhysRevA.90.013409} where the terminal cost is the overlap to the known target state, the ``continuous-time'' variation-principle based quantum computation \cite{Farhi_00, PhysRevLett.103.080502, PhysRevA.90.052317} where the terminal cost is the ground-state energy, and quantum parameter estimation (quantum metrology) \cite{book:Helstrom, book:Holevo, PhysRevLett.96.010401, AdvancesQuantumMetrology_2011, PhysRevX.6.031033, PhysRevA.96.012117, PhysRevA.96.032310, PhysRevA.96.042114, PhysRevA.103.052607} where the cost function is the classical or quantum Fisher information. In a more general context, OCT has been used for the stabilization of ultracold molecules \cite{PhysRevA.70.013402}, optimizing the performance in nuclear magnetic resonance measurement \cite{PhysRevA.90.023411, 10.1063/1.4906751, KOBZAR2012142, PhysRevA.101.012321}, cooling of quantum systems \cite{PhysRevA.82.063422, doi:10.1137/100818431, PhysRevA.87.043607}, charging a two-level quantum battery \cite{PhysRevA.107.032218, PhysRevA.110.052601}, and optimizing the quantum emitter \cite{PRXQuantum.2.040354}.

Quantum two-level system (qubit) is at the heart of many important technologies, such as
Nuclear Magnetic Resonance \cite{book:91190666}, Electron Paramagnetic Resonance spectroscopy \cite{book:EPR-1972} and atomic clock \cite{book:AtomicClock}.
Some applications require manipulating the quantum state. For example, in Magnetic Resonance Imaging or general quantum sensing schemes
one needs to prepare a state that is a superposition of its natural eigenstates \cite{PhysRev.78.695, book:MRI, RevModPhys.92.015004, RevModPhys.89.035002,  Fabricant_2023, 10.1116/5.0025186};
in quantum computation the single-qubit gates are essential for implementing the universal gate set \cite{Neilson_book, book:Kaye_book, 10.1063/1.5089550, RevModPhys.75.281, PhysRevA.57.120, Adams_2020, Bluvstein_2023}.
\change{Qubit states can be manipulated using multiple fields \cite{10.1063/1.4906137, PhysRevA.90.062302, PhysRevA.92.052314, PhysRevA.107.032613} and various} experimental techniques such as AC Stark tuning \cite{10.1063/5.0012257} or dichromatic excitation \cite{PhysRevLett.126.047403}, and in this paper we consider a more common qubit system having only one single scalar control.
When the scalar control \change{is constrained by a maximum amplitude} $u_\text{max}$, OCT classifies the control into bang control and singular control. The former corresponds to $\pm u_\text{max}$ whereas the latter may take any values in between. The intuition brought by OCT (from analyzing linear systems) is that the bang-bang protocol is a strong candidate of time-optimal control \cite{book:GeometricOptimalControl}; when this is indeed the case the parametrization of optimal protocol is greatly simplified. This intuition is the basis of some quantum algorithms, notably QAOA (Quantum Approximate Optimization Algorithm) \cite{Farhi_16, PhysRevX.7.021027}, which \change{relies exclusively on} bang controls to find the ground state.
For generic quantum tasks, the optimal control typically involves a singular component \cite{PhysRevA.100.022327, PhysRevLett.126.070505, PhysRevA.105.042621, PhysRevA.103.052607} \change{whose values are unknown prior to optimization.} For a dissipationless qubit, the allowed singular control is known {\em a priori} \cite{10.1063/1.2203236, PhysRevA.100.022327}, which can be used to constrain the functional space of optimal protocol.

Our work particularly focuses on the influence of the maximum control amplitude and is complementary to the existing results \cite{10.1063/1.2203236, PhysRevA.107.032613, PhysRevA.108.062425} in the following three aspects.
First, the time-optimal protocol of a qubit for the state preparation has been shown to be of multiple bangs \cite{PhysRevA.108.062425, ansel2024introduction} or of bang-singular-bang \cite{10.1063/1.2203236, PhysRevLett.111.260501, PhysRevA.90.032110, PhysRevA.100.022327, PhysRevA.101.022320}. We provide a more complete picture by demonstrating that a sufficiently large control amplitude is the key for the singular control.
We also point out two minor but subtle points about the meta-stable solution and the bang duration which were omitted in the literature.
Second, we apply OCT to the X-gate of a qubit where the global phase matters.
The X-gate is typically accomplished by the resonant Rabi $\pi$-pulse with some minor modifications \cite{PhysRevLett.103.110501, PhysRevA.83.012308, PhysRevA.88.062318}; the derivation is based on Rotating Wave Approximation that neglects the high-frequency components. Our main finding is that the optimal gate time is about 80\%  of the Rabi $\pi$-pulse and approaches $\pi/4$ of the latter in the small amplitude limit. The intuition behind the Rabi protocol is resonance, and we shall see how the resonant behavior and bang-bang protocol reconcile in OCT analysis.
From the physical point of view, the bang-bang solution implies that the high-frequency components and the full calculation are essential for the ultimate quantum speedup.
Finally, to construct a realizable protocol we address the issue of bang-bang control by proposing a few methods that smooth the sharp changes of the time-optimal bang-bang protocol but at the same time preserve the gate fidelity. The smoothed solutions are only meaningful when the evolution time is longer than the minimum gate time using the bang-bang protocol. For this task the bang-bang solution provides a theoretical minimum gate time and serves as a starting point for more realistic considerations.

The paper is organized as follows. In Section II we define the system and the control problem and review the OCT. The features specific to a qubit of unitary dynamics are explicitly pointed out. General optimality conditions used to regulate the optimal protocol are derived; the conditions that rule out singular control are provided.
In Section III we consider a generic state preparation problem. We shall show how the amplitude constraint affects the optimal protocol, particularly the emergence of a singular control.
In Section IV, we apply the OCT to find the minimum time to achieve the X-gate of a qubit. The general control protocol is found to be bang-bang. The minimum gate time is about 20\% shorter than the widely used protocol based on resonant Rabi $\pi$-pulse. The small-amplitude limit is derived and the effects of high frequency are discussed.
In Section V, we provide a few methods to suppress the high-frequency components introduced by the time-optimal bang-bang protocol while maintaining the gate fidelity. 
A brief conclusion is given in Section VI.
In the Appendix \ref{app:planar_dyn} we give the qubit dynamics in terms of angular variables on the Bloch sphere. Appendix \ref{app:odd_freq} provides heuristics on why the odd harmonics play an important role on qubit dynamics based on perturbation.

\section{General features of qubit system}

In this section, we give a short introduction of OCT. While extensive reviews have been presented in the literature \cite{PRXQuantum.2.030203, PRXQuantum.2.010101}, we emphasize its consequences on the qubit system of unitary dynamics. OCT is typically formulated in terms of real-valued dynamical variables, but quantum systems are naturally described by complex-valued wave functions. We shall keep the derivations in the complex-valued form consistent with Schr\"odinger equation that can facilitate generalization to systems of higher dimensions.
We begin with a brief recapitulation of the OCT on generic quantum systems and introduce the optimality conditions and relevant terminologies required for later discussion. The practical usefulness of optimality conditions will be explicated. For a qubit with unitary dynamics, there are important features \change{arising from} its low-dimensionality; there are additional constraints specific to the Hamiltonian and initial/target qubit states. Altogether, the general behavior derived from OCT analysis enables rigorous few-parameter parametrizations of the time-optimal protocol, which will be tested and applied to concrete examples in the following sections.

\subsection{Optimal control on generic quantum systems}  \label{sec:OCT_general}

The quantum system with a single-scalar control can be described by the Hamiltonian
\beq
\begin{aligned}
H(t) &= H_0 + u(t) H_1 
\end{aligned}
\label{eqn:H_general}
\eeq
$H_0$ is the system Hamiltonian, and $H_1$ is the externally applied term whose amplitude $u(t)$ is the scalar that we can control. The optimal control problem for the state preparation is formulated as follows. Given (i) an initial state $| \psi_\text{init} \rangle$, (ii) a target state $| \psi_\text{target} \rangle$, (iii) the maximum control amplitude $u_\text{max}$, and (iv) a total evolution time $T$, find the optimal control $u^*(t)$ that minimizes
\beq
\mathcal{C}^\text{term}_\text{SP}[u(t); T] = -\big| \langle \psi_\text{target}| \mathcal{U}( u(t) ) | \psi_\text{init} \rangle \big|^2
\label{eqn:C_st}
\eeq
where $\mathcal{U}( u(t); T ) = \mathcal{T} e^{-i \int_0^T H(t) \dd t}$ is the evolution operator defined by a control $u(t)$ at a given $T$ and $\mathcal{T}$ denotes time-ordering. Eq.~\eqref{eqn:C_st} is referred to as the ``terminal cost''. For a time-optimal control problem, we are given (i)-(iii) and aim to identify the optimal control $u^*(t)$ and the shortest evolution time $T^*$ that lead to the global minimum of $\mathcal{C}_\text{SP}[u(t); T]$ (-1 in this case). We would like to point out that, in experiments, we typically have direct control over the amplitude of a specific control variable within a given setup, which must be a finite quantity. For instance, in the qubit scenario discussed in this paper, the voltage amplitude generated by the arbitrary waveform generator represents such a variable. Hence, constraint (iii) aligns well with practical considerations.

OCT provides a set of necessary conditions for the optimal solution $u^*(t)$; they can be used to constrain the parametrization of optimal protocol and quantify the quality of a numerical solution.
OCT for the quantum system of Eq.~\eqref{eqn:H_general} are summarized as follows.
\begin{subequations}
\begin{align}
\mathcal{H}_\text{oc} (t) &= \text{Re} \big[ -i \langle \lambda(t) | H(t) | \psi(t) \rangle \big] \nonumber \\
&\change{ = \frac{1}{2i} \bigg[ \langle \lambda(t) | H(t) | \psi(t) \rangle - \langle  \psi(t) | H(t) | \lambda(t) \rangle \bigg] }
\label{eqn:H_oc_0} \\
 \partial_t | \psi \rangle &= 2 \frac{\delta \mathcal{H}_\text{oc} }{ \delta \langle \lambda(t) |  }  = -i H(t) | \psi \rangle \text{ with } | \psi(0) \rangle = | \psi_\text{init} \rangle,
 \label{eqn:forward}  \\
 \partial_t | \lambda \rangle &= -2\frac{\delta \mathcal{H}_\text{oc} }{ \delta \langle \psi(t) |  }  = -i H(t) | \lambda \rangle \text{ with } | \lambda(T) \rangle =  2 \frac{\delta \mathcal{C}^\text{term} }{ \delta \langle \psi(T) |  },
 \label{eqn:adjoint}  \\
 \frac{\delta \mathcal{C}^\text{term} }{ \delta u(t)  } &= \text{Re} \big[ -i \langle \lambda(t) | H_1 | \psi(t) \rangle \big]  \, \dd t \equiv \Phi(t) \dd t, \label{eqn:Phi(t)_0}\\
 \frac{\delta \mathcal{C}^\text{term} }{ \delta T  } &= \mathcal{H}_\text{oc} (T).
\end{align}
\label{eqn:OC_original_gradient}
\end{subequations}
The control-Hamiltonian $\mathcal{H}_\text{oc}(t)$, adjoint field $|\lambda(t) \rangle$, and switching function $\Phi(t)$ in Eq.~\eqref{eqn:OC_original_gradient} are quantities introduced by OCT that are very informative to characterize the system behavior.
Eq.~\eqref{eqn:H_oc_0} defines the control-Hamiltonian $\mathcal{H}_\text{oc} (t)$ which is a real-valued scalar and should not to be confused with a quantum Hamiltonian which is generally a complex-valued matrix. Eq.~\eqref{eqn:forward} is the Schr\"odinger equation for the wave function $| \psi(t) \rangle$; Eq.~\eqref{eqn:adjoint} is the Schr\"odinger-like equation for the adjoint field $| \lambda(t) \rangle$.
Eq.~\eqref{eqn:Phi(t)_0} defines the switching function $\Phi$ that is proportional to the gradient of the user-defined cost function $\mathcal{C}^\text{term}$ with respect to the control. \change{ Practically Eq.~\eqref{eqn:Phi(t)_0} provides the most efficient way, in terms of both memory and speed, to compute $\frac{\delta \mathcal{C}^\text{term} }{ \delta u(t)  }$ and is widely used in the gradient-based optimization algorithm to obtain a numerical solution \cite{KHANEJA2005296}. Eq.~\eqref{eqn:Phi(t)_0} is the core of the procedure introduced in Section \ref{sec:smoothness_promotion} (see Table~\ref{table:pseudo_code} and related discussion).  }

The control system described by Eq.~\eqref{eqn:H_general} is control-affine [linear in $u(t)$] and time-invariant [the time dependence is solely from $u(t)$] which lead to two general optimality conditions. First, the optimal control is $u^*(t) = -u_\text{max} \text{Sgn}[\Phi]$ when $\Phi \neq 0$ where Sgn denotes the sign function; this is referred to as a bang (B) control as the control is at one of its two boundary values. If $\Phi=0$ over a finite time interval, the control is referred to as a singular (S) control. The optimal solution can be expressed as
\beq
u^*(t) = \begin{cases}
          -u_\text{max} \text{Sgn}[\Phi] \text{ if } \Phi \neq 0 \\
          u_\text{sing}(t) \text{  if } \Phi = 0
         \end{cases}
\label{eqn:u^*(t)}
\eeq
The singular control  $u_\text{sing}$ is generally unknown until the numerical calculation is done, but for a unitary qubit the allowed $u_\text{sing}$ can be determined {\em before} the calculation. Second, $\mathcal{H}_\text{oc}(t)$ is a constant for an optimal solution and is the derivative of the terminal cost function with respect to the evolution time $T$; it is negative when $T<T^*$ (i.e., the terminal cost can further decrease upon increasing $T$) and is zero at the time-optimal solution $T=T^*$. These optimality conditions are the first-order necessary conditions and are always checked for a numerical solution. The minimum time satisfying both $\mathcal{H}_\text{oc} (T^*)=0$ and $\mathcal{C}^\text{term}[u^*(t), T^*]=-1$ (global minimum of $\mathcal{C}^\text{term}$) is used to identify the time-optimal solution. In next two subsections we discuss general properties of bang and singular controls for pure qubit systems without quantum decoherence.

\subsection{Qubit and bang-bang control} \label{sec:qubit_BB}

We now apply the OCT to a typical qubit Hamiltonian
\beq
H(t) = \frac{\omega_0}{2} \sigma_z + u(t) \sigma_x.
\label{eqn:H_2L}
\eeq
Here $\omega_0$ is the natural frequency of the system (i.e., the  difference of two eigenenergies without control) and we take $\omega_0 = 2$ for the reminder of the discussion; they correspond to $H_0= \sigma_z$ and $H_1 = \sigma_x$ in Eq.~\eqref{eqn:H_general}. $u(t)$ is the scalar control knob which is bounded by $|u(t)| \leq u_\text{max}$. How the time-optimal protocol changes under different $u_\text{max}$'s is the main focus of this work.

For later discussion we derive the constraint specific to BB control of Eq.~\eqref{eqn:H_2L} \cite{PhysRevA.108.062425, ansel2024introduction}. For the optimal solution, $\mathcal{H}_\text{oc} (t)$ is a constant [see discussion below Eq.~\eqref{eqn:u^*(t)}] and is denoted as $\mathcal{H}_\text{oc} (t) = -|\lambda_0|$.   Eq.~\eqref{eqn:H_oc_0} implies
\beq
\begin{aligned}
-|\lambda_0| & = \text{Re} \big[ -i  \langle \lambda(t) | \sigma_z | \psi(t) \rangle + u ( -i  \langle \lambda(t) | \sigma_x | \psi(t) \rangle ) \big] \\
\Rightarrow \,\,\,   &  \text{Re} \big[ i  \langle \lambda(t) | \sigma_z | \psi(t) \rangle  \big] = u \Phi + |\lambda_0|
\end{aligned}
\eeq
Taking first and second derivatives of $\Phi$ and using the commutation relation of Pauli matrices leads to
\begin{subequations}
\begin{align}
 \dot{\Phi} &= \text{Re} \big[  -i (+i)  \langle \lambda(t) | [\sigma_z + u \sigma_x, \sigma_x] | \psi(t) \rangle \big] = \text{Re} \big[ +i  \langle \lambda(t) | 2 \sigma_y | \psi(t) \rangle \big],  \label{eqn:Phi_dot} \\
 \ddot{\Phi} &=\text{Re} \big[  +i (+i)  \langle \lambda(t) | [\sigma_z + u \sigma_x, 2 \sigma_y] | \psi(t) \rangle \big] \nonumber \\
 &= - \Omega^2 \Phi - 4 u |\lambda_0| \text{ with } \Omega^2 \equiv 4 (1 + u^2). \label{eqn:Phi_ddot}
 \end{align}
\label{eqn:Phi_01}
\end{subequations}
For the BB control [Eq.~\eqref{eqn:u^*(t)}], the switching function $\Phi(t)$ satisfies 
\beq
4  |\lambda_0|\cdot  u_\text{max} \, \text{sgn}[\Phi(t)] =
\ddot{\Phi} + \Omega^2 \Phi.
\label{eqn:Phi_ddot_equaiton}
\eeq
According to Eq.~\eqref{eqn:Phi_ddot_equaiton}, $\Phi(t)$ is periodic and its frequency is determined as follows.
Over the time interval where $\Phi >0$, the formal solution is
\beq
\Phi(t) 
= A \bigg( \sin(\Omega (t-t_0)) + \frac{|\lambda_0|}{|A|}\frac{4 u_\text{max} }{  \Omega^2 } \bigg).
\label{eqn:Phi(t)_rough}
\eeq
with $A>0$ [for $\Phi <0$, take $A=-|A|$ in Eq.~\eqref{eqn:Phi(t)_rough}]. Zeros of $\Phi$ are given by $\sin(\Omega ( t_i - t_0)) = -  \frac{4 u_\text{max} |\lambda_0|}{ A \Omega^2 }$. The time interval between two adjacent zeros of $\Phi(t)$ is given by
\beq
\overline{T} = t_{i+1}-t_i = \frac{\pi}{\Omega} +
\frac{2}{\Omega} \sin^{-1} \left(\frac{4 |\lambda_0|}{A}
\frac{ u_\text{max} }{ \Omega^2 } \right). 
\label{eqn:Tbar_duration}
\eeq
The angular frequency $\omega_\text{eff}$ of the switching function  $\Phi(t)$ is given by
\beq
\omega_\text{eff} \equiv \frac{\pi}{\overline{T} }
= \frac{\Omega}{  1 +  \frac{2}{\pi} \sin^{-1} \big( \frac{4 |\lambda_0|}{A}
\frac{ u_\text{max} }{ \Omega^2 } \big) } \leq \Omega.
\label{eqn:w_eff}
\eeq
The optimal BB control is
\beq
u_\text{bb}^*(t) = -u_\text{max} \text{Sgn} \bigg[ \cos \big( \omega_\text{eff} (t-t_0) \big) \bigg] \text{ for $0 \leq t \leq T$}.
\label{eqn:optimal_bang_bang_form}
\eeq
Eq.~\eqref{eqn:optimal_bang_bang_form} offers a {\em rigorous} two-parameter parametrization of the optimal protocol which greatly reduces the complexity of the numerical optimization. Our derivation preserves the complex-valued form of $| \lambda \rangle$ and $| \psi \rangle$ without introducing any additional real-valued functions and facilitates generalization, otherwise is equivalent to those given in Refs.~\cite{PhysRevA.108.062425, ansel2024introduction}.

Eq.~\eqref{eqn:optimal_bang_bang_form} implies that the optimal bang duration is a constant except the first and the last bangs. The Fourier transform implies only {\em odd} harmonics of $\omega_\text{eff}$ are important, and a connection to the Schr\"odinger equation is provided in Appendix \ref{app:odd_freq}.
One subtlety which was not pointed out previously \cite{10.1063/1.2203236, PhysRevA.108.062425, ansel2024introduction} deserves our attention.
For the time-optimal solution where $| \lambda_0 | = 0$, one might conclude from Eq.~\eqref{eqn:w_eff} that $\omega_\text{eff} = \Omega$. This is {\em not} true: $\Phi_0(t) = A \sin( \Omega(t-t_0) ) $  alone {\em cannot} satisfy Eq.~\eqref{eqn:Phi_ddot_equaiton} because $\ddot{\Phi}_0 + \Omega^2 \Phi_0 = 0$ but $\text{Sgn}[\Phi_0] \neq 0$. What happens is that both $\Phi$ and $\mathcal{H}_\text{oc}$ vanish for the time-optimal BB solution (i.e., both $A$ and $| \lambda_0 |$ approach zero) and $\omega_\text{eff} $ depends on the ratio  $\lim_{|\lambda_0| \rightarrow 0} \frac{|\lambda_0|}{A}$ which is generally non-zero. This subtlety will be numerically verified by examining the optimality conditions stated in Section \ref{sec:OCT_general}. Whether BB is the optimal protocol or not is generally unknown {\em a priori}; for qubit systems we shall give the conditions that guarantees BB as optimal protocol shortly.


\subsection{Qubit and singular control} \label{sec:OCT_planar}

\begin{figure}[ht]
\begin{center}
\includegraphics[width=0.5\textwidth]{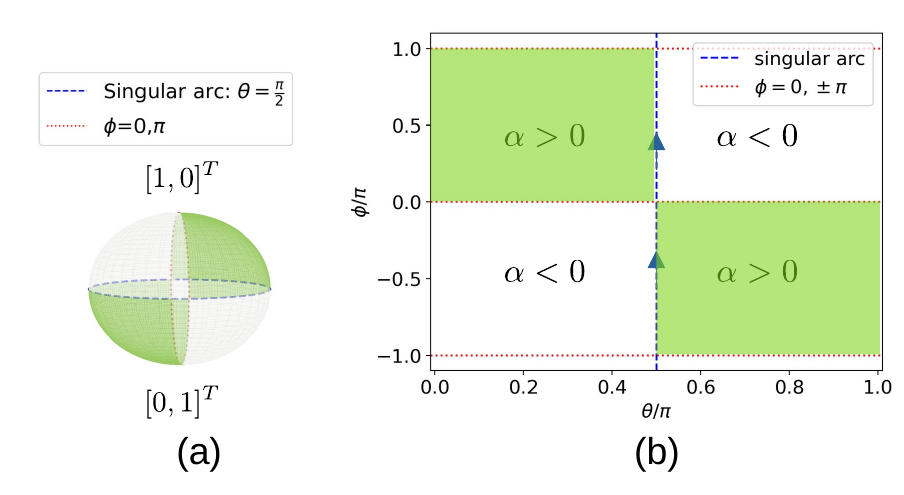}
\caption{Four quadrants on the Bloch sphere defined by the sign of $\alpha$. (a) Bloch sphere. \change{In our convention defined in Eq.~\eqref{eqn:Bloch_representation}, the north pole corresponds to the state $[1,0]^T$ while the south pole $[0,1]^T$. (b) $\theta-\phi$ plane. The green color indicates the quadrant of positive $\alpha$.}  Arrows in (b) indicate the motion of the optimal trajectory along the singular arc.}
\label{fig:alpha=0_independent}
 \end{center}
\end{figure}

A qubit of unitary dynamics is fundamentally a planar system (i.e., two real-valued variables) as a qubit state can be represented by a point on a Bloch sphere or $\theta$-$\phi$ plane:
\beq
| \psi (\theta, \phi) \rangle = \begin{bmatrix} \cos \frac{\theta}{2} \\ \sin \frac{\theta}{2} e^{i \phi} \end{bmatrix}
\Leftrightarrow \mathbf{x} = \begin{bmatrix} \theta \\ \phi \end{bmatrix}
\label{eqn:Bloch_representation}
\eeq
where $\theta \in [0, \pi]$ is the polar angle and $\phi \in (-\pi, \pi]$ the azimuthal. A planar control system has been thoroughly analyzed in Refs.~\cite{Sussmann_87_01, Sussmann_87_02} and is presented in detail in Ref.~\cite{book:GeometricOptimalControl}. The most relevant consequence is that the allowed singular control [$u_\text{sing}$ in Eq.~\eqref{eqn:u^*(t)}] and its corresponding trajectory can be determined {\em without} solving the dynamic equation. We shall sketch the derivation and apply it to Eq.~\eqref{eqn:H_2L}.

The first step is to formulate control dynamics in the planar form $\dot{\mathbf{x}} = \mathbf{f} + u(t) \mathbf{g}$ where $\mathbf{x} = [\theta, \phi]^T$ and $\mathbf{f}$, $\mathbf{g}$ are two real-valued vector fields.
The map between Pauli matrices and their corresponding vector fields are given in Appendix \ref{app:planar_dyn} (see also Ref.~\cite{PhysRevA.100.022327, PhysRevA.101.022320} for detailed derivations). Next we determine a scalar $\alpha(\theta, \phi)$ by $[\mathbf{f}, \mathbf{g}] = \alpha \mathbf{f} + \zeta \mathbf{g}$ where $[\mathbf{f}, \mathbf{g}]$ is the Lie bracket of two vector fields (see Appendix \ref{app:planar_dyn}; $\zeta$ is not needed in the following discussion) \cite{note_computation}.
In the region where $\alpha \neq 0$, the optimal control is BB with at most one switching: $\alpha>0$ only allows the switching from $-u_\text{max}$ to $+u_\text{max}$; $\alpha<0$ only allows the switching from $+u_\text{max}$ to $-u_\text{max}$  (see Chapter 2.9 of Ref.~\cite{book:GeometricOptimalControl}).
$\alpha=0$ defines a ``singular arc'' which is the only trajectory that allows a singular control. Moreover the allowed singular control $u^*_\text{sing}$ is shown to be a Hamiltonian-dependent constant for a qubit of unitary dynamics \cite{PhysRevA.101.022320}.

For the qubit system specified by Eq.~\eqref{eqn:H_2L}, $u^*_\text{sing}=0$ and $\alpha = \frac{ 2 \cot \theta }{\sin \phi}$ [see Appendix \ref{app:planar_dyn}]. The singular arc $\alpha=0$ corresponds to $\theta=\frac{\pi}{2}$, the equator of Bloch sphere [Fig.~\ref{fig:alpha=0_independent}(a)]. $\alpha$ changes sign when crossing the singular arc and the arcs specified by $\phi=0, \pm \pi$. In the $\theta$-$\phi$ plane, the sign of $\alpha$ defines four quadrants as illustrated in Fig.~\ref{fig:alpha=0_independent}(b).
If the optimal protocol involves a singular part, during that period $u^*(t)$ has to be zero (i.e., $H=\sigma_z$) and the qubit has to stay on the singular arc where only $\phi$ is increasing (subject to a modulo of $2\pi$) whereas  $\theta$ is fixed at $\frac{\pi}{2}$. This implies that if the initial and/or target state is at one of two poles of Bloch sphere where $\phi$ is irrelevant for the state specification, the time-optimal control {\em cannot} involve the singular part. According to Eq.~\eqref{eqn:u^*(t)}, $u^*(t)$ for these problems has to be BB which can be parametrized by Eq.~\eqref{eqn:optimal_bang_bang_form}. Although not generic, qubit states at two poles are crucial in quantum gates and will be discussed in Section \ref{sec:Xgate-qubit}.



\subsection{Summary of OCT analysis} \label{sec:C1-3}

Three general constraints from OCT for the unitary qubit system described by Eq.~\eqref{eqn:H_2L} are summarized below.
\bi
\item (C1) If the optimal control is BB and involves two switchings or more, it can be rigorously parameterized by the two-parameter form of Eq.~\eqref{eqn:optimal_bang_bang_form}. In particular the durations of all middle bangs are identical.

\item (C2) Sign of $\alpha$ divides the Bloch sphere into four quadrants of alternate signs [Fig.~\ref{fig:alpha=0_independent}]. When the optimal trajectory is in the region of the same sign of $\alpha (\neq 0)$, the optimal control is BB with at most one switching.

\item (C3) The optimal trajectory of singular control can only happen along $\alpha=0$ [$\theta = \frac{\pi}{2}$ for Eq.~\eqref{eqn:H_2L}], which is referred to as the singular arc. When it happens, $u^*_\text{sing}=0$ and the trajectory satisfies $\dot{\phi}= 2$.

Because along the singular arc only the azimuthal $\phi$ is changing, S control cannot be part of time-optimal protocol if the initial and/or target states are at the poles of Bloch sphere where $\phi$ is irrelevant for state specification; in these cases the optimal protocol is BB.

\ei

To significantly change the quantum state with a weak $u_\text{max}$, the oscillation frequency $u(t)$ has to match the natural frequency $\omega_0$. This is known as the resonance condition (see for example Ref.~\cite{book:quantum_optics}). The oscillatory behavior is encoded in constraints (C1) and (C2) as the BB protocol is only consistent with the trajectory that swings between quadrants of opposite signs.  When $u_\text{max}$ is comparable to the natural frequency, a non-resonant solution including the singular control can emerge and is constrained by the constraint(C3). In the next two sections we provide examples to illustrate these behaviors.

Combining constraints (C2) and (C3), we conclude that the optimal control $u^*(t)$ has to be piece-wise constant with $u^* = \pm u_\text{max}$ for the bang control or $u^*=0$ for the singular control. This allows us to parametrize $u^*(t)$ using a set of $u$ values and switching times. Consider an optimal control composed of $\{ u_i\}_{i=1}^N = (u_1, u_2, \cdots, u_{N} )$ ($u_i \in [0, \pm u_\text{max}]$) segments of $N-1$ switchings at $(t_1, t_2, \cdots, t_{N-1})$. Denoting $t_0=0$ and $t_N=T$, the total evolution operator is given by
\beq
\mathcal{U}( u(t) ) = \mathcal{U}( \{ t_i \}_{i=1}^{N-1} ; \{ u_i \}_{i=1}^{N} ) = \Pi_{i=1}^{N-1}  U(t_{i+1} - t_{i}, u_i).
\label{eqn:u^*_parametrization}
\eeq
where $\Pi_{i=1}^{N}O_i = O_N \cdots O_2 O_1$ and  $U(t,u)$ is the evolution operator for a constant $u$:
\beq
U(t,u) = e^{ -i (\sigma_z + u \sigma_x) t }
= \cos( \Omega_0(u) \, t ) \sigma_0 - i \sin( \Omega_0(u) \, t )  \frac{\sigma_z + u \sigma_x}{\Omega_0(u) },
\label{eqn:U(t,u)}
\eeq
with $\Omega_0(u) = \sqrt{1+u^2}$. The number of switching $N-1$ is roughly determined by the resonance condition: $N-1 \approx \frac{T}{\pi/\omega_0} = \frac{\omega_0 T}{\pi}$. 
Once the control system is beyond planar [such as the damped qubit or multiple qubits], the value of singular control cannot be pre-determined anymore
but the condition of vanishing $\ddot{\Phi}$ can still be used to construct the optimal control involving the singular part \cite{PhysRevLett.126.070505, PhysRevA.105.042621, PhysRevA.101.022320}.

%

\section{State preparation}

In this section we analyze a state preparation problem. The main purpose is to examine how $u_\text{max}$ affects the optimal protocol.
In particular we show that there exists a critical $u_c$ above which the time-optimal control allows a singular part. Similar problems have been investigated in Refs.~\cite{10.1063/1.2203236, PhysRevA.90.032110} \change{and more recently in the context of charging quantum battery \cite{PhysRevA.110.052601}}, and we provide a sharper picture by (i) showing the existence of a meta-stable solution around  $u_c$; and (ii) examining the subtle relationship between the bang duration and Eqs.~\eqref{eqn:Phi_ddot_equaiton} and \eqref{eqn:Phi(t)_rough}.

\subsection{Overview and problem statement}

For the generic state preparation, the terminal cost is given by Eq.~\eqref{eqn:C_st}. For a concrete problem we choose the initial and target states to be $| \psi_\text{init} \rangle =  | \psi(\theta_\text{init}, \phi_\text{init}) \rangle$,
$| \psi_\text{target} \rangle =  | \psi(\theta_\text{target}, \phi_\text{target}) \rangle$ with $(\theta_\text{init}, \phi_\text{init}) = ( 0.7 \pi , 0)$, $(\theta_\text{target}, \phi_\text{target}) = ( 0.35 \pi, \pi)$ [see Eq.~\eqref{eqn:Bloch_representation}]. This choice is very close to the problem considered in Ref.~\cite{PhysRevA.101.022320}: the rationale is that two states are sufficiently far from each other to allow for the potential non-trivial control.
We shall vary $u_\text{max}$ to obtain the time-optimal solution and the optimal time $T^*(u_\text{max})$.

The optimization procedure is done as follows.
To minimize Eq.~\eqref{eqn:C_st} we first choose a set of $\{ u_i \}$ in Eq.~\eqref{eqn:u^*_parametrization} and use switching times as independent variables, i.e.,
\beq
\mathcal{C}_\text{SP}^\text{term} ( \{ t_i \}_{i=1}^{N-1} ; \{ u_i \}_{i=1}^{N}  )
= -\big| \langle \psi_\text{target} | \mathcal{U}( \{ t_i \}_{i=1}^{N-1} ; \{ u_i \}_{i=1}^{N} ) | \psi_\text{init} \rangle \big|^2.
\label{eqn:C_switchings}
\eeq
For a given $\{ u_i \}$, we first minimize Eq.~\eqref{eqn:C_switchings} using Nelder-Mead algorithm \cite{10.1007/s10589-010-9329-3} to obtain switching times $\{ t_i \}_{i=1}^{N-1}$ and then  apply two optimality conditions ($u^*(t) = - u_\text{max} \text{Sgn}[\Phi(t)]$ and constant $\mathcal{H}_\text{oc}(t)$) to examine if the resulting control is a local optimum.
The similar procedure has been used previously in Refs.~\cite{PhysRevA.100.022327, PhysRevA.101.022320}. The time-optimal solution is obtained by gradually increasing the total evolution time $T$ until both $\mathcal{H}_\text{oc}(T^*) = 0$ and $\mathcal{C}_\text{SP}^\text{term}=-1$ are satisfied.

\subsection{Results and discussion}
\begin{figure}[ht]
\begin{center}
\includegraphics[width=0.8\textwidth]{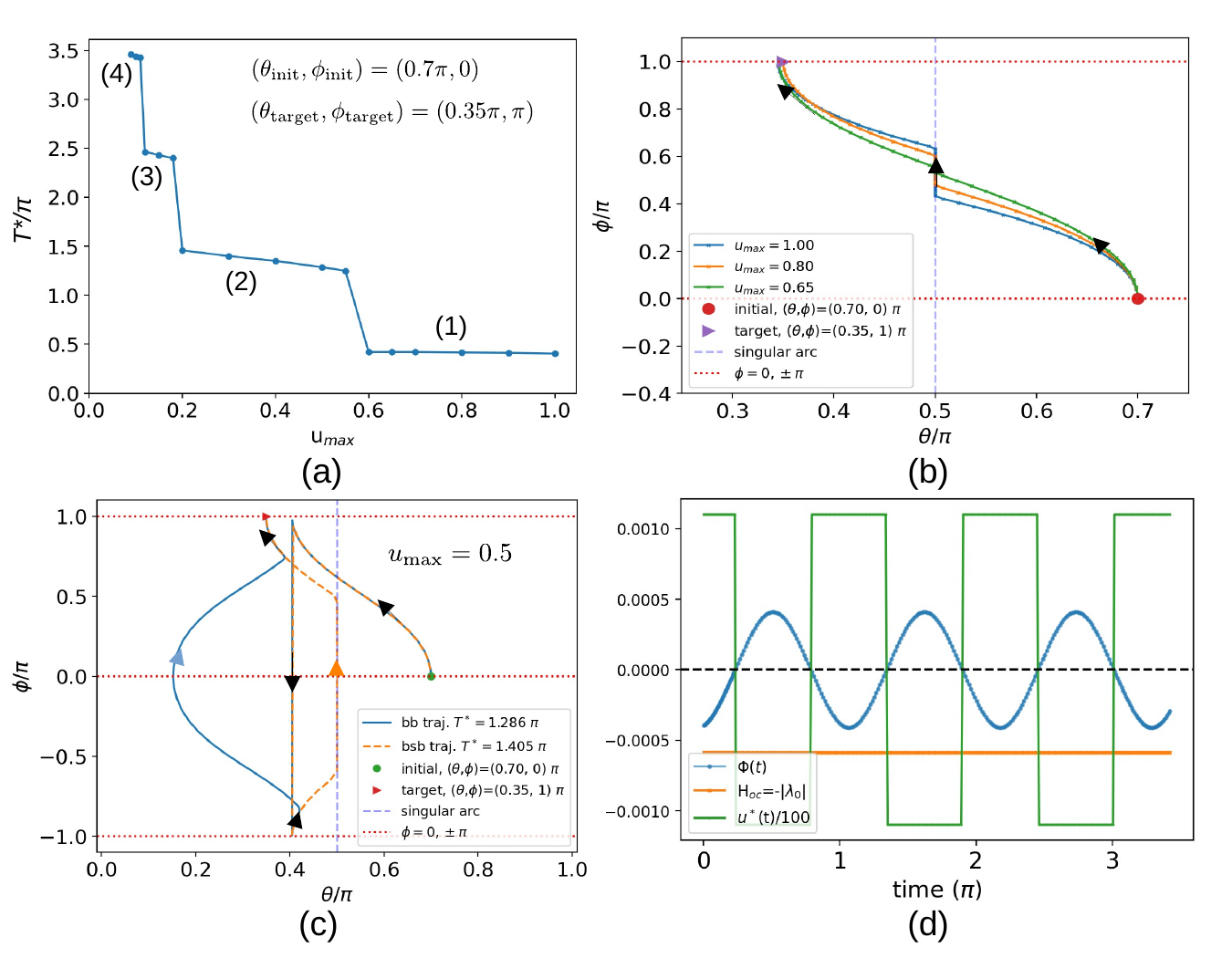}
\caption{ (a) The optimal time $T^*$ as a function of maximum amplitude $u_\text{max}$.
Four plateaus of $T^*$ appear in this range and they correspond to (1) BSB, (2) BB-2, (3) BB-4, (4) BB-6 where the number denotes the number of switchings. (b) The optimal trajectories for BSB. The segment on singular arc decreases upon reducing $u_\text{max}$. (c) For $u_\text{max}=0.5$, the optimal protocol is BB-2. Both BSB and BB-2 are local minima and their trajectories are shown. Arrows in (b) and (c) indicate the direction of the trajectory. Notice that $\phi = \pm \pi$ are the same point, \change{and at the vertical segment where $\theta \approx 0.4\pi$, $\phi$ goes from $\pi$ to $-\pi$ in BB-2 trajectory instantaneously.} (d) The optimal control for $u_\text{max}=0.11$ where $T^*=$3.4285 $\pi$; both optimality conditions ($u^*(t) = -u_\text{max} \text{Sgn}[\Phi(t)]$ and a constant $\mathcal{H}_\text{oc}(t)$) are examined. The duration of middle bangs is about 0.56 $\pi$. 
}
\label{fig:T^*_state_preparation}
 \end{center}
\end{figure}

The main result is summarized in Fig.~\ref{fig:T^*_state_preparation}(a) which shows $T^*(u_\text{max})$ and the structures of the corresponding optimal controls.
As expected, the larger the maximum amplitude $u_\text{max}$ the shorter the optimal time $T^*$. For $u_\text{max} \in [0.1, 1]$ there are four plateaus of $T^*$; they are labeled as (1) to (4) and correspond to controls of different number of switchings.
When $u_\text{max} > u_c \sim 0.6$ [plateau (1)], the optimal control is BSB. Three BSB optimal trajectories are shown in Fig.~\ref{fig:T^*_state_preparation}(b). At the singular control the trajectory indeed stays on the singular arc defined by $\theta=\pi/2$ [see the discussion in Section \ref{sec:OCT_planar} and (C3) in Section \ref{sec:C1-3}].
Upon reducing $u_\text{max}$ the portion of optimal trajectory on the singular arc becomes shorter and eventually disappears at $u_\text{max} = u_c$. We point out that $u_c$ depends on initial and target states: if we change the initial state to $(\theta_\text{init}, \phi_\text{init}) = ( 0.65 \pi , 0)$, $u_c = 0.51$. From Fig.~\ref{fig:T^*_state_preparation}(b) it is clear that $u_c$ happens when the trajectories starting from initial and target points under one of allowed bang controls (can be $\pm u_\text{max}$) intersect at the singular arc. \change{
It is worth noting that the optimal BSB control found in Ref.~\cite{PhysRevA.110.052601} (an effective qubit system with dynamics similar to Eq.~\eqref{eqn:H_2L} in the context of quantum battery) shares a great similarity with our analysis in two aspects. First Ref.~\cite{PhysRevA.110.052601} considers the system of large $u_\text{max}$ ($\frac{ u_\text{max} }{\omega_0} \geq 1.25$ in our convention) which is larger than $u_c$ and could favor the S control. Second the singular portion of optimal trajectory also decreases upon reducing $u_\text{max}$. The key difference is that in Ref.~\cite{PhysRevA.110.052601} the global phase of the effective qubit state is relevant in the original system and cannot be neglected . The effective qubit in Ref.~\cite{PhysRevA.110.052601} is therefore not planar anymore but involves three real-valued variables; for this reason the $u^*_\text{sing}$ and the singular arc derived in Section \ref{sec:OCT_planar} cannot be directly applied. }

Once $u_\text{max} < u_c \sim 0.6$ the optimal control are found to be BB with different number of switchings.
Only BB with even number of switchings are found to be optimal; this is {\em not} general but specific to the choice of initial/target states.
\change{We use BB-$n$ to indicate the BB controls with $n$ number of switchings.} When $0.2 <  u_\text{max} < 0.6$ [plateau (2)], the optimal control is BB with two switchings (BB-2). The optimal trajectory of $u_\text{max} = 0.5$ is given in Fig.~\ref{fig:T^*_state_preparation}(c). Around $u_\text{max} \lesssim u_c$, the BSB solution is still a local minimum and its trajectory is also shown Fig.~\ref{fig:T^*_state_preparation}(c). The time-optimal solution is obtained by picking the one with a shorter gate time. We are not aware of a rigorous selection rule, but empirically the time-optimal trajectory tends to avoid staying on the singular arc for too long as the singular control does not (fully) utilize the control to move the state variables.

The plateaus of $T^*$ reflect the number of switchings of BB controls. There is a jump in $T^*$ when the number of switchings increases; within the same number of switchings $T^*$ only varies gradually.
Fig.~\ref{fig:T^*_state_preparation}(d) shows the optimal control $u^*(t)$ for $u_\text{max}=0.11$ that has 6 switchings. Both switching function $\Phi(t)$ and control-Hamiltonian $\mathcal{H}_\text{oc}$ are plotted to illustrate the numerical accuracy to which the optimal conditions are satisfied. Two important features related to constraint (C1) in Section \ref{sec:C1-3} are highlighted. First, the time durations of middle bangs are identical which is consistent with (C1). The duration is numerically found to be around $0.56 \pi$, unambiguously larger than $\frac{ \pi}{ 2(1+u^2_\text{max}) } \approx 0.497 \pi$ obtained by simply taking $|\lambda_0 | (=-\mathcal{H}_\text{oc}) = 0$ in Eq.~\eqref{eqn:Tbar_duration}. Second, upon approaching the time-optimal solution $T \rightarrow T^*$, both $|\mathcal{H}_\text{oc} |$ and $\Phi$ vanish, consistent with the discussion below Eq.~\eqref{eqn:optimal_bang_bang_form}.

Overall, when $u_\text{max}$ exceeds a state-dependent critical value, a shortcut can emerge and the optimal protocol includes a singular control. In the other limit where $u_\text{max}$ is small compared to the natural frequency $\omega_0 (=2)$, a fidelity one state preparation requires a control of multiple BB which resembles a resonant behavior. 
The state preparation between generic initial/target states can be relevant in some effective qubit systems \cite{PhysRevLett.79.325, PhysRevA.57.2403, PhysRevA.100.022327, PhysRevA.110.052601}.
For a physical qubit system, the initial/target states are typically easily prepared states or Hamiltonian-specific eigenstates; this will be considered in Section \ref{sec:Xgate-qubit}.

\section{X-gate of qubit} \label{sec:Xgate-qubit}

\subsection{Overview}

In this section we apply OCT to the X-gate of qubit. \change{Compared to the state preparation, the qubit gate operation is more complicated in that the global phase matters, and this additional requirement is reflected in the terminal cost function.} Following  Ref.~\cite{PhysRevLett.103.110501} the cost function of X-gate operation is chosen to be
\beq
\mathcal{C}_\text{X} [ u(t); T ] = -\frac{1}{4} \big| \langle 1 | \mathcal{U}(u(t)) |0 \rangle
+ \langle 0 | \mathcal{U}(u(t)) |1 \rangle \big|^2
\label{eqn:cost_xgate}
\eeq
Here $| 0 \rangle \equiv [1,0]^T$ and $| 1 \rangle \equiv [0,1]^T$ are two eigenstates of $H_0 = \sigma_z$. At the global minimum, Eq.~\eqref{eqn:cost_xgate} demands the phase from $| 0 \rangle \rightarrow |1 \rangle$ is identical to that from $| 1 \rangle \rightarrow | 0 \rangle$;  the coefficient $\frac{1}{4}$ is chosen such that the global minimum of  $\mathcal{C}_\text{X}$ is -1. We shall use $\mathcal{C}_\text{X}+1$, whose global minimum is zero, to characterize the gate performance.

\begin{figure}[ht]
\begin{center}
\includegraphics[width=0.45\textwidth]{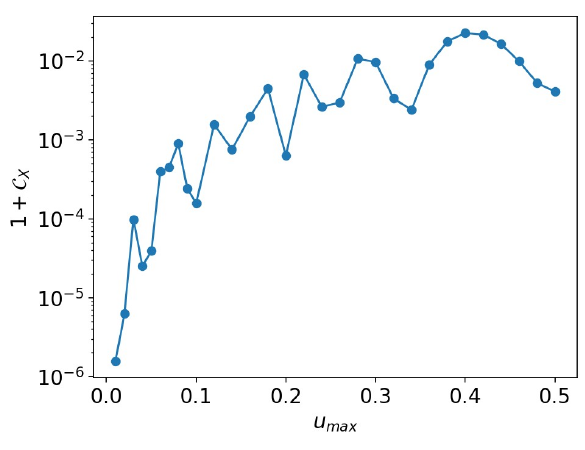}
\caption{ $1+\mathcal{C}_X$ as a function of $u_\text{max}$ using $u^\text{Rabi}(t)$ given Eq.~\eqref{eqn:Rabi_pi}. $1+\mathcal{C}_X$ approaches zero only when $u_\text{max}$ is small as the resonant Rabi protocol is based on RWA but $1+\mathcal{C}_X$ is computed without any approximation.
}
\label{fig:Fidelity_Rabi}
 \end{center}
\end{figure}

The widely used resonant Rabi $\pi$-pulse \cite{PhysRevLett.103.110501} is used as the reference:
\beq
u^\text{Rabi}(t) = u_\text{max} \cos\big( \omega_0 (t-\frac{T}{2}) \big) \text{ for } 0\leq t \leq T=T^\text{Rabi}_{\pi} = \frac{\pi}{u_\text{max}}.
\label{eqn:Rabi_pi}
\eeq
$u^\text{Rabi}(t)$ is even with respect to $t=T/2$ as required by the X-gate (see next subsection for a detailed explanation). Its gate time $T^\text{Rabi}_{\pi} $ will serve as the baseline for comparison.
The resonant Rabi oscillation is based on the Rotating Wave Approximation (RWA) that neglects the high-frequency components and is valid when $\omega_0 \gg u_\text{max}$. We point out that using Eq.~\eqref{eqn:Rabi_pi} within RWA, the X-gate is of fidelity one, but in the full calculation there is a small deviation due to the high-frequency components and therefore the gate is never complete. To explicitly show this we compute $\mathcal{C}_X+1$ (zero when the gate is complete) for $u_\text{max} \in [0.01, 0.5]$ using the ``resonant Rabi protocol'' Eq.~\eqref{eqn:Rabi_pi}. As shown in Fig.~\ref{fig:Fidelity_Rabi}, $\mathcal{C}_X+1$ approaches zero only when $u_\text{max} \rightarrow 0$.
The resonant Rabi protocol only involves the resonant frequency during the evolution. We shall see that the time-optimal solution utilizes the high-frequency components even in the $\omega_0 \gg u_\text{max}$ limit and can reduce the total gate time by 20\% with respect to $T^\text{Rabi}_{\pi}$.

\subsection{Parametrization of control and optimization}

Given the cost function Eq.~\eqref{eqn:cost_xgate}, one can deduce a few conditions that further constrain the optimal control. First, because the initial/target states have no $\phi$ dependence  $u^*$  is expected to be strictly BB [see the discussion (C3) in Section \ref{sec:OCT_planar}]. Using constraint (C1), the optimal evolution operator at final time $T$ is BB of equal middle-bang duration: 
\beq
\mathcal{U}_\text{X,bb}(T) = U( \tilde{t}_1, (-1)^N u_\text{max})
\bigg[ \Pi_{i=2}^{N-1} U( \tilde{t}, (-1)^i u_\text{max})  \bigg] U( \tilde{t}_0, -u_\text{max})
\label{eqn:U_x_bb}
\eeq
with $\tilde{t}_0 + \tilde{t}_1 + (N-2) \tilde{t} = T$, $\tilde{t}_0, \tilde{t}_1 < \tilde{t}$, and $U(t,u)$ given by Eq.~\eqref{eqn:U(t,u)}.
Knowing BB being the time-optimal protocol, two general features can be derived.
$\mathcal{U}_\text{X,bb}(T)$ being an ideal X-gate implies $\mathcal{U}_\text{X,bb}(T) = e^{i \phi} \sigma_x$ (arbitrary $\phi$) and $\sigma_z \mathcal{U}_\text{X,bb}(T) \sigma_z$ is also a perfect X-gate as
\beq
\begin{aligned}
\mathcal{U}_\text{X,bb}(T) = e^{i \phi} \sigma_x
\Rightarrow \sigma_z \mathcal{U}_\text{X,bb}(T) \sigma_z
= -e^{i \phi} \sigma_x = e^{i (\phi+\pi)} \sigma_x.
\end{aligned}
\eeq
Using $\sigma_z \sigma_z \sigma_z = \sigma_z$ and $\sigma_z \sigma_x \sigma_z = -\sigma_x$, we get $\sigma_z U(t,u) \sigma_z = U(t,-u)$ and
\beq
\begin{aligned}
\sigma_z \mathcal{U}_\text{X,bb}(T) \sigma_z & = \sigma_z U( \tilde{t}_1, (-1)^N u_\text{max}) \sigma_z
\bigg[ \Pi_{i=2}^{N-1} \sigma_z U( \tilde{t}, (-1)^i u_\text{max}) \sigma_z \bigg]
\sigma_z U( \tilde{t}_0, -u_\text{max}) \sigma_z \\
&=U( \tilde{t}_1, (-1)^{N+1} u_\text{max})
\bigg[ \Pi_{i=2}^{N-1} U( \tilde{t}, (-1)^{i+1} u_\text{max})  \bigg] U( \tilde{t}_0, +u_\text{max}).
\end{aligned}
\label{eqn:U_x_bb_sigma_z}
\eeq
Since Eqs.~\eqref{eqn:U_x_bb} and \eqref{eqn:U_x_bb_sigma_z} correspond to BB protocols of opposite signs, $\pm u^*(t)$ are degenerate optimal solutions.
Also, $\mathcal{U}_\text{X,bb}$ equals to its transpose $(\mathcal{U}_\text{X,bb} )^T$. Using $U(t,u) = U^T(t,u)$ one gets
\beq
\big( \mathcal{U}_\text{X,bb} \big)^\text{t} = U( \tilde{t}_0, -u_\text{max})
\bigg[ \Pi_{i=N-1}^{2} U( \tilde{t}, (-1)^i u_\text{max})  \bigg]
 U( \tilde{t}_1, (-1)^N u_\text{max}).
\label{eqn:U_x_bb_T}
\eeq
Equating Eq.~\eqref{eqn:U_x_bb_T} with Eq.~\eqref{eqn:U_x_bb}, we conclude $N$ is an odd integer and $\tilde{t}_1= \tilde{t}_0$ (the first and last bangs have the same time durations). These two conditions are equivalent to $u^*(t-\frac{T}{2}) = + u^*(\frac{T}{2}-t)$, i.e., the optimal protocol is even with respect to $t = \frac{T}{2}$.

The analysis above indicates that the optimal protocol can be parametrized rigorously by
\beq
\begin{aligned}
u^*(t; \omega_\text{eff}(T, u_\text{max}) ) &= \pm u_\text{max} \times \text{Sgn}
\bigg[   \cos\bigg( \omega_\text{eff}(T, u_\text{max}) \cdot (t-\frac{T}{2}) \bigg) \bigg].
\end{aligned}
\label{eqn:u*_Xgate_1p}
\eeq
This form, depending only on one single parameter $\omega_\text{eff}(T, u_\text{max}) $, is valid for any $T\leq T^*(u_\text{max})$. 
The optimization is done by expressing Eq.~\eqref{eqn:cost_xgate} as a function of  $\omega_\text{eff}$; because it is now a one-dimension problem one can use bisection method to find the global minimum. With Eq.~\eqref{eqn:u*_Xgate_1p}, the duration of middle bangs is given by $\overline{T} = \frac{\pi}{ \omega_\text{eff} }$ which is determined numerically.
From the discussion in Section \ref{sec:OCT_general}, the switching function $\Phi(t)$ can be parametrized analytically as
\beq
\Phi(t) =
\begin{cases}
A \cos \big(\Omega (t-\frac{T}{2}) \big) + \frac{4 u_\text{max} |\lambda_0|}{ \Omega^2 }, &
(2n-\frac{1}{2})\overline{T} < t-\frac{T}{2} < (2n+\frac{1}{2})\overline{T} \\
A \cos \big(\Omega (t-\frac{T}{2}) \big) {-} \frac{4 u_\text{max} |\lambda_0|}{ \Omega^2 }, &
(2n+\frac{1}{2})\overline{T} < t-\frac{T}{2} < (2n+\frac{3}{2})\overline{T}
\end{cases}
\label{eqn:Phi_analytical}
\eeq
Here $A>0$, $n$ is any integer, and $0\leq t \leq T$. Notice that $\overline{T}$ and $\frac{|\lambda_0|}{A}$ are {\em not} independent but related by Eq.~\eqref{eqn:Tbar_duration}; the latter is non-zero when $|\lambda_0| \rightarrow 0$.

For completeness we mention that the cost functions for the Y-gate and for the population transfer between $| 0 \rangle$/$|1 \rangle$ can be chosen as
\beq
\begin{aligned}
\mathcal{C}_\text{Y} [ u(t); T ] &= -\frac{1}{4} \big| \langle 1 | \mathcal{U}(u(t)) |0 \rangle - \langle 0 | \mathcal{U}(u(t)) |1 \rangle \big|^2, \\
\mathcal{C}_\text{PT} [ u(t); T ] &= -\frac{1}{2} \bigg( \big| \langle 1 | \mathcal{U}(u(t)) |0 \rangle \big|^2 + \big| \langle 0 | \mathcal{U}(u(t)) |1 \rangle \big|^2 \bigg).
\end{aligned}
\label{eqn:cost_PT}
\eeq
For the Y-gate, the rigorous one-parameter parametrization is obtained by replacing $\cos \big( \omega_\text{eff}(t-T/2) \big)$ by $\sin \big( \omega_\text{eff}(t-T/2) \big)$ in Eq.~\eqref{eqn:u*_Xgate_1p} so that $u^*$ is odd with respect to $t=\frac{T}{2}$ [i.e., $u^*(t-\frac{T}{2}) = -u^*(\frac{T}{2}-t)$]. This conclusion is obtained by recognizing $\mathcal{U}_\text{Y,bb} = e^{i \phi} \sigma_y$ and utilizing $( \mathcal{U}_\text{Y,bb} )^T = -\mathcal{U}_\text{Y,bb} = + \sigma_x  \mathcal{U}_\text{Y,bb} \sigma_x$.
For population transfer, the global phase does not matter so both forms are legitimate and have to be considered.

\subsection{Time-optimal solution}

\begin{figure}[ht]
\begin{center}
\includegraphics[width=0.8\textwidth]{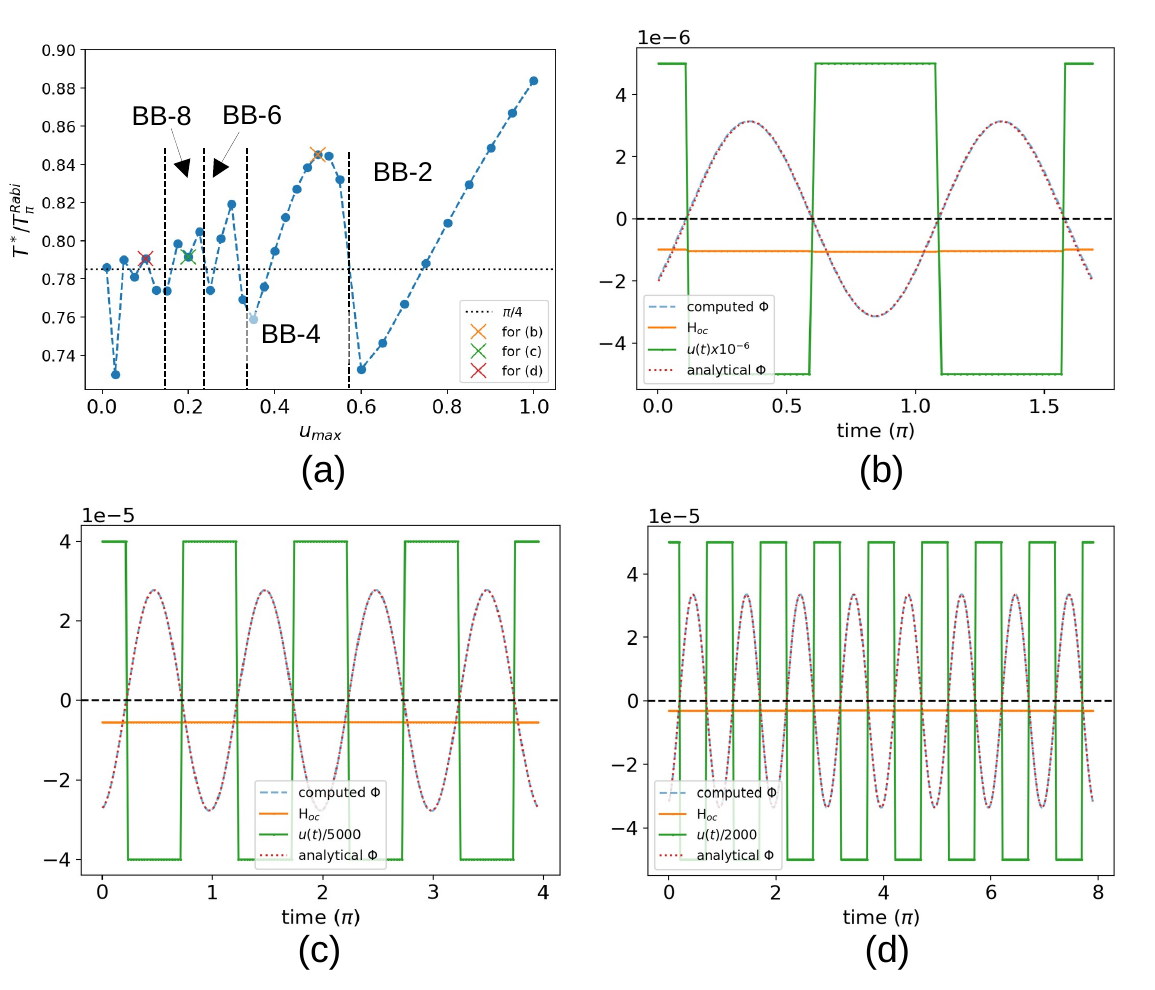}
\caption{ (a) The ratio between the optimal-time and the Rabi $\pi$-pulse $T^*/T^\text{Rabi}_\pi$ as a function of maximum amplitude $u_\text{max}$; this ratio approaches $\pi/4$ when $u_\text{max} \rightarrow 0$. 
Boundaries between BB-2, BB-4, BB-6, BB-8 are indicated as vertical dashed lines. Only even number of switchings are allowed for the X-gate because $u^*(t)$ is even with respect to $t=T/2$.  Time-optimal solutions for the three crosses are given in (b), (c), (d).
(b) $u_\text{max}=0.5$ where $u^*$ is BB-4 and $\omega_\text{eff} = 2.0435$.
(c) $u_\text{max}=0.2$ where BB-8 and $\omega_\text{eff} = 1.9899$.
(d) $u_\text{max}=0.1$ where BB-16 and $\omega_\text{eff} = 1.9979$.
In (b)-(d), two optimality conditions, a constant $\mathcal{H}_\text{oc}(t) = -|\lambda_0|$ and $\Phi$ and $u^*(t)$ being opposite in sign, are well satisfied. We purposely choose a gate time slightly shorter than $T^*$ so that $|\lambda_0|$ is small but non-zero \cite{note_gatetime}.
}
\label{fig:T^*_Xgate}
 \end{center}
\end{figure}

In this subsection we present the time-optimal solutions for different values of $u_\text{max}$, emphasizing (i) the time reduction with respect to the Rabi protocol and (ii) the optimality of the one-parameter control Eq.~\eqref{eqn:u*_Xgate_1p}.  Fig.~\ref{fig:T^*_Xgate}(a) summarizes the overall behavior:
the ratio between the optimal gate time and Rabi $\pi$-pulse $T^*/T^\text{Rabi}_{\pi}$
is about 0.8 and approaches $\pi/4$ in the $u_\text{max}\rightarrow 0$ limit (proven in next subsection); the optimal protocol is BB with an increasing number of bangs upon reducing $u_\text{max}$.
To quantitatively examine the optimal solution,
$u^*(t)$ for $u_\text{max}$ =0.5, 0.2, 0.1 at $T \lessapprox T^*(u_\text{max} )$ are respectively given in Fig.~\ref{fig:T^*_Xgate}(b), (c), (d) \cite{note_gatetime}.
The optimal control does obey  Eq.~\eqref{eqn:u*_Xgate_1p} as the optimality conditions given in Section \ref{sec:OCT_general} are well satisfied. Numerical optimization gives $\omega_\text{eff}=2.0435$ for $u_\text{max}=0.5$; $\omega_\text{eff}=1.9899$ for $u_\text{max}=0.2$; $\omega_\text{eff}=1.9979$ for $u_\text{max}=0.1$.
These values are clearly smaller than their respective $\Omega = 2 \sqrt{1 + u_\text{max}^2}$ which are respectively 2.236 ($u_\text{max}=0.5$), 2.040 ($u_\text{max}=0.2$) and 2.010 ($u_\text{max}=0.1$).
We further verify the switching function Eq.~\eqref{eqn:Phi_analytical} by determining $A$ from Eq.~\eqref{eqn:w_eff} with $|\lambda_0|$ and $\omega_\text{eff}$ obtained from the optimal solution. Substituting $A$ and $\lambda_0$ into Eq.~\eqref{eqn:Phi_analytical}  results in a switching function that agrees with that obtained by a direct evaluation using Eq.~\eqref{eqn:Phi(t)_0} [Fig.~\ref{fig:T^*_Xgate}(b)-(d)].

\subsection{Small $u_\text{max}$ limit}


We now consider $T^*/T^\text{Rabi}_\pi$ in the $u_\text{max} \rightarrow 0$ limit which serves two purposes: first it grants an analytical expression;  second it is the limit where RWA becomes exact so one can convincingly see the necessity of the full calculation for the time-optimal control.
In this limit both $T^*$ and $T^\text{Rabi}_\pi$ diverge but their ratio can be obtained by counting the number of oscillations over the gate time. Given the natural frequency $\omega_0 (=2)$, one oscillation in $u(t)$ takes $t_0 = 2 \pi/ \omega_0 (=\pi)$. This period holds for the Rabi $\pi$-pulse with arbitrary $u_\text{max}$ values, but for the BB protocol {\em only} when $u_\text{max} \rightarrow 0$. Over the duration of $ T^\text{Rabi}_\pi = \frac{\pi}{ u_\text{max} } $, the number of oscillations is $N_\text{Rabi} = \frac{ T^\text{Rabi}_\pi }{ t_0  } = \frac{\omega_0}{2 u_\text{max} } = \frac{1}{ u_\text{max} }$.

To obtain the number of oscillation for the BB protocol $N_\text{bb}$, we consider the following evolution operator over an oscillation period $t_0$:
\beq
U_\text{bbX} = U( \frac{t_0}{4}, u_\text{max} ) U(\frac{t_0}{2},-u_\text{max} ) U( \frac{t_0}{4}, u_\text{max} ). 
 \label{eqn:u_bbX}
\eeq
$(U_\text{bbX})^N$ represents a BB protocol of the evolution time $N t_0$. The order of Eq.~\eqref{eqn:u_bbX} ensures that (i) the durations of all middle bangs of $(U_\text{bbX})^N$ are identically $t_0/2$ as required by constraint (C1), and (ii) $(U_\text{bbX})^N$ is even with respect to the middle evolution time as required by X-gate.
Using Eq.~\eqref{eqn:U(t,u)} and keeping only the linear order in $u_\text{max}$, Eq.~\eqref{eqn:u_bbX} is reduced to
\beq
U_\text{bbX} \approx  -\sigma_0 + 2 i u_\text{max} \sigma_x \underset{ u_\text{max} \rightarrow 0 }{\longrightarrow}
-e^{-i 2 u_\text{max} \sigma_x}.
\eeq
Requiring $(U_\text{bbX})^{ N_\text{bb} } \sim e^{-i 2 u_\text{max} N_\text{bb} \sigma_x} = \cos( 2 u_\text{max} N_\text{bb} ) \sigma_0 - i \sin (2 u_\text{max} N_\text{bb}) \sigma_x  \sim \sigma_x$ up to a phase factor, we get $2 u_\text{max} N_\text{bb} = \frac{\pi}{2}$ so that $N_\text{bb} = \frac{\pi}{4} \frac{1}{ u_\text{max} } = \frac{\pi}{4} N_\text{Rabi}$. The asymptotic ratio is determined by
\beq
\frac{ T^* }{ T^\text{Rabi}_\pi } = \frac{ N_\text{bb} t_0 }{ N_\text{Rabi} t_0 }
\underset{  u_\text{max} \rightarrow 0 }{ \longrightarrow }
\frac{\pi}{4}. 
\label{eqn:ratio_asymptotic}
\eeq
The same ratio is obtained for the Y-gate by considering $U_\text{bbY} = U( \frac{t_0}{2}, u_\text{max} ) U(\frac{t_0}{2},-u_\text{max} ) \sim e^{-i 2 u_\text{max} \sigma_y}$ and then requiring $(U_\text{bbY})^{N_\text{bb}} \sim \sigma_y$.

The constant in Eq.~\eqref{eqn:ratio_asymptotic} being independent of model parameters indicates that $T^\text{Rabi}_\pi$ is also an intrinsic characteristic time scale. Indeed $T^\text{Rabi}_\pi$ is the shortest gate time using a single (resonant) frequency in $u_\text{max} \rightarrow 0$ limit. The small $u_\text{max}$ calculation also highlights the importance of the full calculation. As RWA ignores the high frequency components, it can never obtain the true quantum limit even in the $u_\text{max} \rightarrow 0$ limit.

To sum up, we establish that BB control is the time-optimal protocol for X-gate and the obtained $T^*( u_\text{max})$ is the shortest gate time given the amplitude constraint $u_\text{max}$. This result cannot be obtained using RWA as the high frequency components are essential in the optimal control. When $u_\text{max}$'s are small, the obtained $\omega_\text{eff}$ are close to $\omega_0=2$, which is consistent with the resonance behavior. However, BB protocol requires sudden jumps in control at specific times which may not be realistic, and in Section \ref{sec:relaxing} we consider a few methods to relax this requirement.


\section{ Smoothing bang-bang protocol } \label{sec:relaxing}

\subsection{Overview} \label{sec:smooth_overview}

We have established that BB is the time-optimal control for X-gate, and obtain the theoretical minimum gate time $T^*( u_\text{max} )$ for a given amplitude constraint $u_\text{max}$.
In practice, however, the control amplitude is not the only relevant constraint. In this Section we address one apparent issue of BB protocol, namely, the high-frequency components caused by the discontinuities of BB control. We shall consider a few schemes to smooth BB protocol while maintaining the perfect gate fidelity (i.e., $C_\text{X} +1=0$) so that the resulting protocol becomes more feasible.

Let us first describe the role of gate time $T$.
When $T<T^*( u_\text{max} )$, $C_\text{X} +1$ is always greater than zero because the evolution time is too short to complete the gate. At $T = T^*( u_\text{max} )$, BB is the only protocol that can complete the gate. When $T > T^*( u_\text{max} )$, there are infinitely many solutions that satisfy $C_\text{X} +1=0$, and it is in this time regime that one can  promote the solution based on additional criteria without compromising the gate fidelity.

We propose three methods to smooth the BB protocol for $T > T^*( u_\text{max} )$.
The first approach is to smooth the discontinuity in BB protocol in time domain;
\change{the second to shorten the gate time of the resonant Rabi protocol by introducing higher-frequency components.}
Numerically, these two approaches are based on optimization in the restricted functional space; 
they are less computationally expensive but also harder to generalize.
The third one is to define a cost function that quantifies the smoothness and minimizes it. This method requires solving a constrained optimization problem; it is more computationally costly but can be straightforwardly generalized to other realistic considerations.

\subsection{Smoothing BB protocol in time-domain} \label{sec:smooth_FT}

\begin{figure}[ht]
\begin{center}
\includegraphics[width=0.95\textwidth]{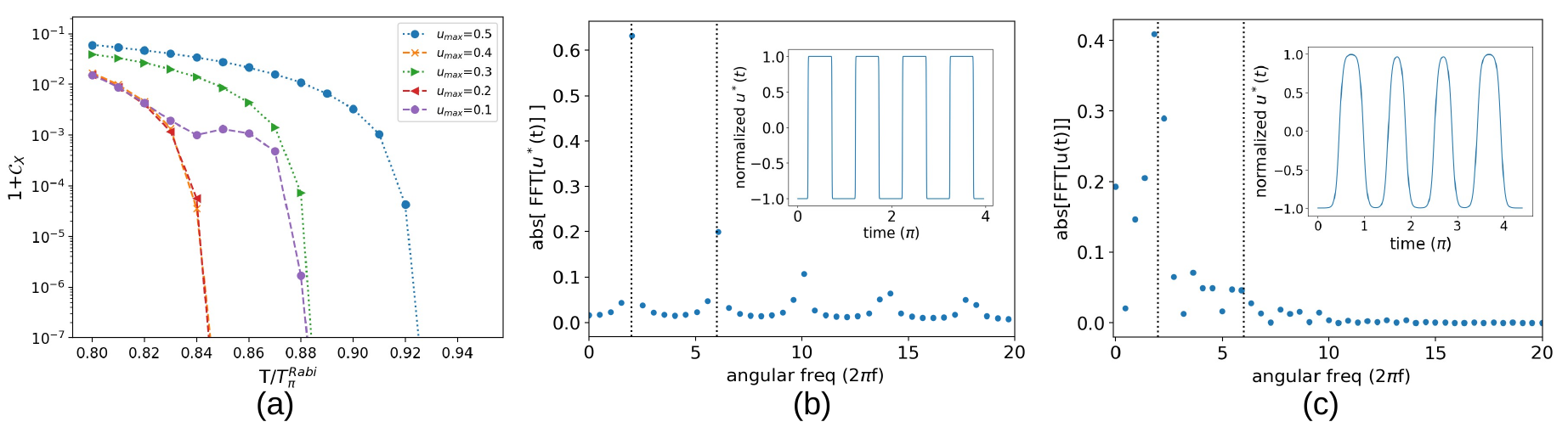}
\caption{
(a) $1+ \mathcal{C}_\text{X} (>0)$ as a function of normalized time $T/T^\text{Rabi}_{\pi}$ with the smoothing parameter $\beta=4$. $u_\text{max}$ = 0.1 to 0.5 are shown.
For $u_\text{max}$ = \change{ 0.1}, the cusp around $T/T^\text{Rabi}_{\pi} = 0.84$ comes from an increase in the number switchings.
(b) Fourier transform of $u^*(t)/u_\text{max}$ with $u_\text{max}=0.2$ [Fig.~\ref{fig:T^*_Xgate}(c)]. $T=T^*\approx 3.958 \pi$.
(c) Fourier transform for the optimal smoothed  BB control where $\beta=4$, $T/T^\text{Rabi}_\pi = 0.88$ (\change{$T=4.4 \pi$}).
Insets of (b), (c) provide the wave form of normalized control $u(t)/u_\text{max}$.
}
\label{fig:FT_smoothing}
 \end{center}
\end{figure}

Our first approach is to replace step function $\Theta(t)$ in BB control by hyperbolic tangent function $\tanh(\beta t)$. Specifically the control protocol is
\beq
\begin{aligned}
u(t; \{ t_i \}_1^N, \beta) &= u_\text{max} \bigg[  \sum_{i=1}^{2N} (-1)^i
\tanh( \beta (t - t_i) ) - 1 \bigg]  \\
\text{ with } & t_i = T-t_{2N+1-i}.
\end{aligned}
\label{eqn:tanh_u}
\eeq
$\beta$ is the smoothing parameter characterizing the smoothness of the switching: smaller $\beta$ corresponds to the smoother switching. The constraint $t_i = T-t_{2N+1-i}$ ensures $u(t-\frac{T}{2}) = u(\frac{T}{2}-t)$. The number of switchings $2N$ is fixed by the resonance condition $2N \approx \frac{T}{\pi/\omega_0} \approx \frac{2 T}{\pi}$; $(t_1, t_2, \cdots, t_N)$ and the first $N$ switching times are independent variables for optimization. The optimization is done using the Nelder-Mead algorithm.

The quantity $1+\mathcal{C}_\text{X} $ (zero for the perfect gate) using the smoothing parameter $\beta=4$ as a function of $T/T^\text{Rabi}_{\pi}$ for $u_\text{max}=$ 0.1 to 0.5 are shown in Fig.~\ref{fig:FT_smoothing}(a). The minimum times to achieve the perfect gate are all shorter than the Rabi $\pi$-pulse; the change of  $T/T^\text{Rabi}_{\pi}$ with respect to $u_\text{max}$ is not monotonic in our calculation. Fig.~\ref{fig:FT_smoothing}(b) and (c) show the Fourier transforms of the normalized pure and smoothed BB control protocol at $u_\text{max}=0.2$. The convention of Fourier transform is chosen as
\beq
\tilde{u}(f_n) =  \int_0^T \dd t\, \frac{ u(t) }{ u_\text{max} } e^{-i 2 \pi f_n t} \text{ where } f_n \equiv \frac{n}{T}.
\eeq
From Fig.~\ref{fig:FT_smoothing}(b), one observes that only odd harmonics of $\omega_0$ have significant contributions. This is a direct consequence of constraint (C1), and a  heuristic argument based on perturbation calculation is given in Appendix \ref{app:odd_freq}. With $\beta=4$, a perfect gate can be achieved when $T/T^\text{Rabi}_{\pi} \approx 0.88$ where the frequency components higher than $5 \omega_0$ are strongly suppressed.

\subsection{ Including the third harmonic }

\begin{figure}[ht]
\begin{center}
\includegraphics[width=0.8\textwidth]{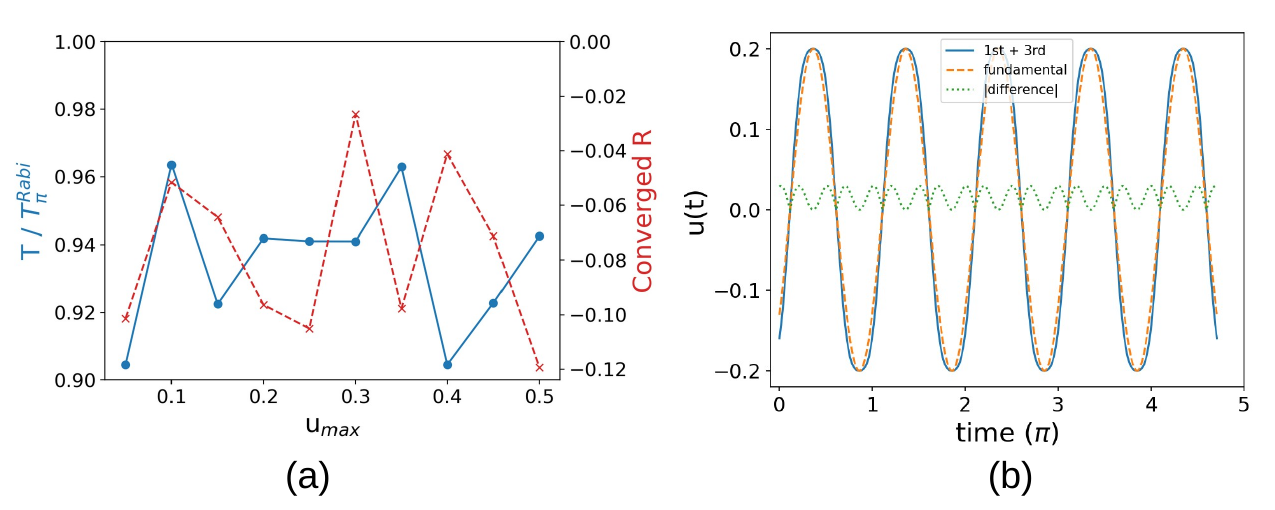}
\caption{ Protocol using only fundamental and third harmonics [Eq.~\eqref{eqn:u_3rd_harmonic}].
(a) The gate time $T$ in $T^\text{Rabi}_\pi$  (blue, left axis) and the optimized ratio $R$ (red, right axis) as a function of $u_\text{max} \in [0.05, 0.5]$.
(b) The corresponding optimized control for $u_\text{max} = 0.2$ (blue solid curve). The control using only the fundamental frequency, i.e., $u(t) = u_\text{max} \cos( \omega (t-\frac{T}{2}) )$ (orange dashed curve) is provided as the reference; their difference in absolute value is given by the green dotted curve.
}
\label{fig:Xgate_harmonic}
 \end{center}
\end{figure}

As discussed in Appendix \ref{app:odd_freq}, only odd harmonics of the fundamental frequency have significant contributions to the evolution when $u_\text{max}$ is small, we thus consider the protocol that contains first and third harmonic with proper symmetry:
\beq
u(t; \omega, R) \equiv u_\text{max} \bigg[  (1 - R) \cos\big( \omega (t-\frac{T}{2} ) \big)
+ R \cos\big( 3 \omega (t-\frac{T}{2} ) \big)   \bigg], \text{ where }-\frac{1}{8} \leq  R \leq 1.
\label{eqn:u_3rd_harmonic}
\eeq
The constraint of $R$ ensures that the maximum amplitude is $u_\text{max}$. We minimize $C_\text{X}$ with respect to both $R$ and $\omega$ at given $T$ and $u_\text{max}$, and then scan $T$ to find the evolution time such that $C_\text{X} + 1 =0$. The optimization over $R$ and $\omega$ is done using Nelder-Mead algorithm. As shown in Fig.~\ref{fig:Xgate_harmonic}(a), the time reduction is about 5-10\% when including the third harmonic. For $u_\text{max} \in [0.05, 0.5]$, all optimized $\omega$'s are very close to $\omega_0=2$ (not shown); all optimized $R$ are negative [right $y$-axis in Fig.~\ref{fig:Xgate_harmonic}(a)], resulting in a flatter profile around the extrema of $u(t)$ compared to that using single frequency. As an illustration, the controls with and without third harmonic, and their difference, for $u_\text{max}=0.2$ are given in Fig.~\ref{fig:Xgate_harmonic}(b). One can see that the third harmonic indeed brings the control closer to BB.

\subsection{ Promoting the smoothness by optimization } \label{sec:smoothness_promotion}

\begin{figure}[ht]
\begin{center}
\includegraphics[width=0.8\textwidth]{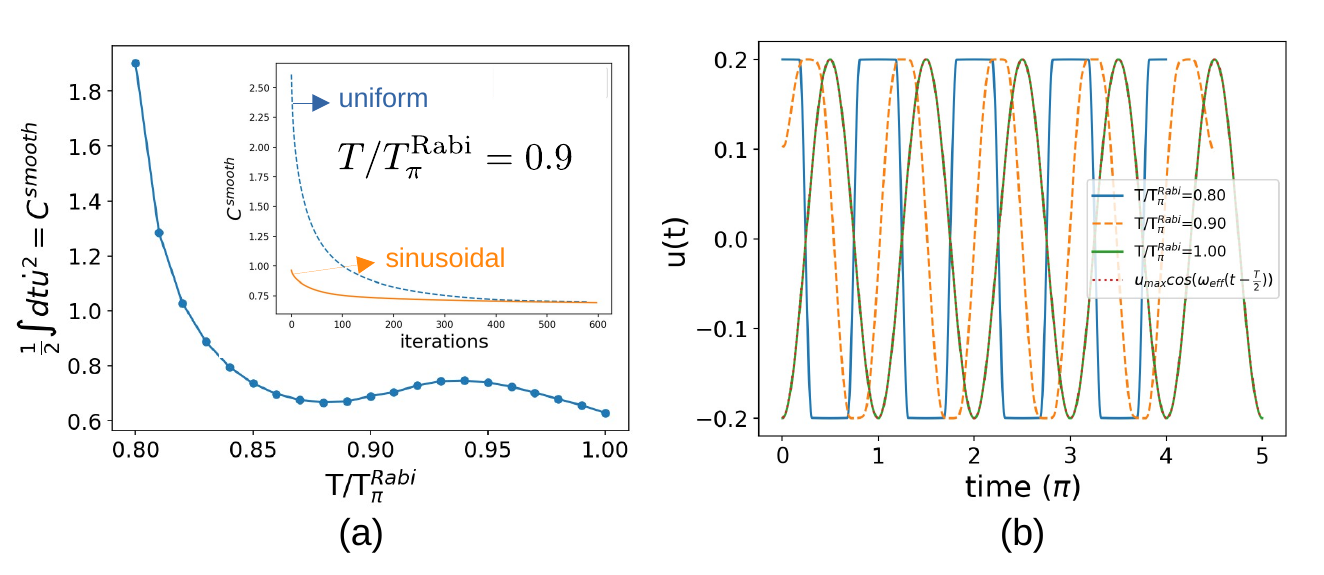}
\caption{ Promoting the smooth solution by solving Eq.~\eqref{eqn:smoothness_cost}. Results of $u_\text{max} = 0.2$ using $N_t=1000$ [Eq.~\eqref{eqn:u(t)_discrete}] are shown.
(a) The converged $\mathcal{C}^\text{smooth} [ u(t) ]$ for $T/T^\text{Rabi}_\pi \in [0.8, 1.0]$. The inset shows the convergence for $T/T^\text{Rabi}_\pi$ = 0.9 starting from two different initial protocols.
(b) The resulting controls for $T/T^\text{Rabi}_\pi$ = 0.8, 0.9, 1.0 \change{following the procedure outlined in Table \ref{table:pseudo_code}}. At $T=T^\text{Rabi}_\pi = 5 \pi$, the converged control matches $u(t) = u_\text{max} \cos( \omega_\text{eff} (t-T/2) )$ ($\omega_\text{eff} \approx 1.995$ here).
}
\label{fig:smoothness_optimization}
 \end{center}
\end{figure}

Our third approach to promote the smoothness is by solving a constrained optimization problem. The starting point is to select a cost function that quantifies the smoothness; we use $\mathcal{C}^\text{smooth} [ u(t) ] = \frac{1}{2} \int_0^T \dd t \, \dot{u}^2(t) $ which is zero when $u(t)$ is a constant. Its gradient is $\frac{\delta \mathcal{C}^\text{smooth} }{ \delta u(t) } = - \dd t \, \ddot{u}$ if we choose the boundary condition such that $u(t) \dot{u}(t) |_{t=0,T} = 0$; we adopt the Neumann boundary condition $\dot{u}(0) = \dot{u}(T) = 0$. Because the gate fidelity cannot be compromised, we consider a {\em constrained} optimization problem
\beq
\underset{ \text{ s.t. } \mathcal{C}_X [u(t)] = -1, | u(t) | \leq u_\text{max} }{ \text{min} \, \mathcal{C}^\text{smooth} [u(t)]  }.
\label{eqn:smoothness_cost}
\eeq
The constraint ensures a perfect gate at a given $u_\text{max}$. As mentioned in Section \ref{sec:smooth_overview}, this problem is only meaningful for the evolution time $T>T^*$: among the infinitely many $u$'s that satisfy $\mathcal{C}_\text{X} [u(t)] +1=0, | u(t) | \leq u_\text{max}$, the optimization promotes the one that minimizes $\mathcal{C}^\text{smooth} [u(t)]$.
To numerically solve Eq.~\eqref{eqn:smoothness_cost}, $u(t)$ is parametrized by a piece-wise constant function
\beq
u(t;\{ u_i \}_{i=1}^{N_t} ) \equiv \sum_{i=1}^{N_t} u_i \Theta(t - (i-1) \Delta t ) \Theta( i\, \Delta t - t) \text{ with } \Delta t = \frac{T}{N_t}.
\label{eqn:u(t)_discrete}
\eeq
Using OCT, we have a gradient-based algorithm \cite{KHANEJA2005296, PhysRevA.101.022320} that efficiently solves $\mathcal{C}_\text{X} [u]=-1$.
In the $T>T^*$ regime, the solution depends on the initial $u_\text{init}$ and we use the subscript of $u_{ u_\text{init} }(t) $ to indicate that the solution of $\mathcal{C}_\text{X} [u_{ u_\text{init} }(t)] =-1$ is obtained from the initial control $u_\text{init}$.

\begin{table}[ht]
 \begin{tabular}{cl}
  Step & Description \\ \hline
  1 & Start with an arbitrary control $\tilde{u}^{(0)}(t)$ \\
     & update ${u}^{(n)}(t)$, $\tilde{u}^{(n)}(t)$ $(n \neq 1)$ by the following procedure: \\
  2a & Update $u^{(n)}(t)$ by solving $C_\text{X}[ u^{(n)} ] = -1$ using a gradient-based algorithm  \\
     & with the initial guess of $\tilde{u}^{(n-1)}(t)$; gradient $\frac{\delta \mathcal{C}_X}{\delta u(t)}$ is computed using  Eq.~\eqref{eqn:Phi(t)_0}. \\
  2b &  Update $\tilde{u}^{(n)}$ using $\tilde{u}^{(n)} = u^{(n)}_{ \tilde{u}^{(n-1)} } - \dd u^{(n)} \, \frac{\delta \mathcal{C}^\text{smooth} }{ \delta u(t) }$, where \\
  & $\dd u^{(n)} $ is chosen such that $\text{max}\big[ \dd u^{(n)} \, \frac{\delta \mathcal{C}^\text{smooth} }{ \delta u(t) } \big] = \frac{u_\text{max} }{5}$ \\
  3. & Repeat steps 2a, 2b until $| u^{(n+1)}(t) - u^{(n)}(t) | \leq \varepsilon \equiv \frac{u_\text{max}}{4000}$ for all $t$. \\ 
  4. & Output the converged $u(t)$.
 \end{tabular}
 \caption{ \blue{ Step-by-step description of the iterative procedure to solve Eq.~\eqref{eqn:smoothness_cost}.} }
\label{table:pseudo_code}
\end{table}

We now describe the iterative procedure to solve Eq.~\eqref{eqn:smoothness_cost}.
At each iteration we introduce $u^{(n)}$ and $\tilde{u}^{(n)}$; the former (without tilde) satisfies the equality constraint whereas the latter (with tilde) does not.
At $n$th iteration, ${u}^{(n-1)}$, $\tilde{u}^{(n-1)}$ are updated by 
\begin{subequations}
\begin{align}
 & C_\text{X}[ u^{(n)}_{ \tilde{u}^{(n-1)} } ] = -1, \label{eqn:equality_u^n} \\
 & \tilde{u}^{(n)} = u^{(n)}_{ \tilde{u}^{(n-1)} } - \dd u^{(n)} \, \frac{\delta \mathcal{C}^\text{smooth} }{ \delta u(t) } . \label{eqn:minimizing_tilde_u} 
\end{align}
\label{eqn:iterative_02}
\end{subequations}
In Eq.~\eqref{eqn:iterative_02}, step \eqref{eqn:equality_u^n} imposes the equality constraint and thus gate fidelity whereas \eqref{eqn:minimizing_tilde_u} aims to minimize $\mathcal{C}^\text{smooth}$. \change{In Eq.~\eqref{eqn:equality_u^n},  $u^{(n)}_{ \tilde{u}^{(n-1)} } $ is updated by minimizing $\mathcal{C}_\text{X}[u(t)]$ using a gradient-based algorithm with $\tilde{u}^{(n-1)}$ as the initial control; $\frac{\delta \mathcal{C}_X}{\delta u(t)}$ is computed using  Eq.~\eqref{eqn:Phi(t)_0}}. In Eq.~\eqref{eqn:minimizing_tilde_u}, $\tilde{u}^{(n)}$ is updated by moving $u^{(n)}_{ \tilde{u}^{(n-1)} }$ along the $-\frac{\delta \mathcal{C}^\text{smooth} }{ \delta u(t) }$ to minimize $\mathcal{C}^\text{smooth}$; the stepsize $\dd u^{(n)}$ is chosen so that the maximum amplitude of  $\dd u^{(n)} \, \frac{\delta \mathcal{C}^\text{smooth} }{ \delta u(t) }$ equals to $\frac{u_\text{max} }{5}$.
The iteration starts with an initial $\tilde{u}^{(0)}$, and continues until $| u^{(n+1)} - u^{(n)} |$ is smaller than a tolerance (chosen to be $\frac{u_\text{max} }{4000}$). Once converged, the optimized control is $u^{(n)}(t)$, not $\tilde{u}^{(n)}$, because $u^{(n)}(t)$ is the one that satisfies the equality constraint and maintains the gate fidelity.
\change{The complete procedure is summarized in Table \ref{table:pseudo_code}. }
The solution from the proposed procedure can depend on the hyper-parameters of the iterative solver (mainly the updating rates), but our numerical tests show the differences are very small for the converged results. It also turns out that Sequential Least Squares Programming algorithm (SLSQP) \cite{SLSQP}, a version of sequential quadratic programming \cite{NoceWrig06} implemented in SciPy (an open-source Python library), gives very similar results.

Results of $u_\text{max}=0.2$, whose $T^* \approx 0.78 T^\text{Rabi}_\pi$, are shown in Fig.~\ref{fig:smoothness_optimization}. The simulations are done using $N_t = 1000$ in Eq.~\eqref{eqn:u(t)_discrete} and we have tested the convergence upon increasing $N_t$.
The converged  $\mathcal{C}^\text{smooth}$ for $T/T^\text{Rabi}_\pi \in [0.8, 1.0]$ is given in Fig.~\ref{fig:smoothness_optimization}(a). The inset shows how $\mathcal{C}^\text{smooth}$ decreases upon iterating Eq.~\eqref{eqn:iterative_02} [step 2 in Table~\ref{table:pseudo_code}] starting from two different initial $u(t)$'s for $T/T^\text{Rabi}_\pi$ = 0.9. Fig.~\ref{fig:smoothness_optimization}(b) shows the resulting controls for $T/T^\text{Rabi}_\pi$ = 0.8, 0.9, 1.0. It is clearly seen that the longer gate time $T$ leads to the smoother protocol. At $T=T^\text{Rabi}_\pi$, the converged protocol well matches $u(t) = u_\text{max} \cos( \omega_\text{eff} (t-T/2) )$ ($\omega_\text{eff} \approx 1.995$), indicating the single-frequency Rabi protocol minimizes the smoothness cost $\mathcal{C}^\text{smooth}$. Once $T > T^\text{Rabi}_\pi = \pi/u_\text{max}$, the amplitude constraint becomes inactive as the gate can now be complete with a smaller maximum amplitude.

The formalism provided here can be directly applied to any user-defined cost function.  For example, one can use $\mathcal{C}^\text{power} = \frac{1}{2}\int \dd t \, u^2$ if the goal is to minimize the total power consumption, or a weighted sum over $\mathcal{C}^\text{power}$ and $\mathcal{C}^\text{smooth}$ if both smoothness and power consumption are important. Typically the cost function is chosen to have nice properties such as being differentiable and/or convex. From the practical point of view, OCT provides an efficient evaluation of the gradient with respect to the cost function which makes the optimization over the free-form parametrization [Eq.~\eqref{eqn:u(t)_discrete}] feasible.  

\section{Conclusion}

We apply OCT to analyze the time-optimal control of pure qubit systems,  focusing particularly on \change{the impact of } the amplitude constraint $u_\text{max}$ on the time-optimal solution.
OCT \change{proves to be highly} effective in constructing the time-optimal protocol for qubit because the underlying dynamics is control-affine and time-invariant, and moreover the qubit only has two real-valued degrees of freedom.
By utilizing the general optimality conditions and constraints specific to the planar control system, the optimal control $u^*(t)$ is shown to be piece-wise constant with 0 (singular control) and $\pm u_\text{max}$ (bang control) being the only three permitted values. These constraints allow us to use switching times as the independent variables to parametrize the optimal protocol which greatly reduces the degrees of freedom for optimization.

Two classes of problems have been considered.
We first consider the generic state preparation problem, where the objective is to guide the qubit from the initial state to the target state in the shortest time. We find that there exists a state-dependent critical amplitude $u_c$ above which the singular control emerges. Below $u_c$ the optimal protocol is BB with the number of switchings increases upon decreasing $u_\text{max}$.
We then consider the X-gate of a qubit, where the objective is to complete a {\em state-independent} operation in the shortest time. For this task the global phase \change{plays a crucial role} in the sense that the X-gate requires the global phase generated upon steering $|0\rangle$ to $|1 \rangle$ to be identical to that $|1\rangle$ to $|0 \rangle$.
The time-optimal protocol is found to be BB; no singular control is allowed. Moreover,
based on some symmetry considerations the time-optimal protocol can be rigorously parameterized by a single-variable form. The minimum gate time $T^*$ is approximately 20\% shorter than the widely used  Rabi $\pi$-pulse $T^\text{Rabi}_\pi$; as the sinusoidal \change{waveform of the Rabi protocol is replaced by square pulses.}. When the maximum control amplitude approaches zero, $\frac{T^*}{T^\text{Rabi}_\pi} \rightarrow \frac{\pi}{4}$. 
\change{Since BB protocols contain abrupt discontinuities that may be challenging to implement experimentally,}
three exemplary methods are provided to suppress the high-frequency components while maintaining perfect gate fidelity so that the resulting protocol is more plausible.
\change{Considering these factors}, the gate time $T$ must be longer than $T^*$, or the gate cannot be completed.
The first method is based on smoothing the BB protocol, and the second is based on
adding third harmonic component to the Rabi $\pi$-pulse. These two methods are relatively easy to implement but hard to generalize.
The third method is based on a constrained optimization, where the objective is the control smoothness and the constraint is the perfect gate operation. \change{We developed and numerically tested a procedure to solve this problem, demonstrating its flexibility in optimizing other user-defined features, such as low power consumption.}
\change{Finally, we emphasize that the time-optimal solution extends beyond the Rotating Wave Approximation (RWA).} \change{Even in the limit} $u_\text{max} \ll \omega_0$, \change{obtaining the time-optimal solution requires a full calculation, indicating that} the high-frequency components are essential for accelerating quantum tasks.

\section*{Acknowledgment}
C. L. thanks Arvind Raghunathan for very helpful discussions on the optimization procedure. Q. D. is grateful to William D. Oliver and Jeffery A. Grover for their support and helpful discussions.
\appendix

\section{Control field on Bloch sphere} \label{app:planar_dyn}

To analyze the dynamics in $(\theta, \phi)$ manifold, we need to map the Hamiltonian in the Schr\"odinger equation to a vector field. Any $2 \times 2 $ Hermitian matrix is a linear combination of three Pauli matrices and the identity matrix with the corresponding vector fields \cite{PhysRevA.100.022327}:
\beq
\begin{aligned}
\sigma_z &\rightarrow V_z = 2 \partial_\phi
\\ 
\sigma_x &\rightarrow V_x = -2 \sin\phi \, \partial_\theta
- 2 \cos \phi \cot\theta \,  \partial_\phi
\\ 
\sigma_y &\rightarrow V_y = 2 \cos\phi \, \partial_\theta
-2 \sin \phi  \cot \theta \, \partial_\phi.
\end{aligned}
\label{eqn:vector_pauli}
\eeq
$\{\partial_\theta, \partial_\phi\}$ is the basis on the tangent space of $(\theta, \phi)$ manifold. Note that the following commutation relations hold: $[V_z, V_x] = -2 V_y$, $[V_y, V_z] = -2 V_x$, and $[V_x, V_y] = -2 V_z$. The commutator of two vector fields $\mathtt{a}= a_i \partial_i$ and $\mathtt{b}= b_i \partial_i$, known as Lie bracket, is given by $[\mathtt{a}, \mathtt{b}] = a_i \partial_i ( b_j \partial_j ) - b_i \partial_i ( a_j \partial_j )$ where repeated indices are summed over.
The identity matrix that generates a global phase has no effect on the dynamical variables $(\theta, \phi)$.

The equation of motion on Bloch sphere for Eq.~\eqref{eqn:H_2L} with a constant $u(t) = u$ is
\beq
\begin{bmatrix} \dot{\theta} \\ \dot{\phi} \end{bmatrix} = V_z + u(t) V_x
= \begin{bmatrix} -2 u \sin \phi \\ 2 - 2 u \cos \phi \cot \theta \end{bmatrix}
\eeq
Define $[V_z,V_x] = \alpha(\theta,\phi) V_z + \beta(\theta,\phi) V_x = -2 V_y$, we get
\beq
-2 V_y = \begin{bmatrix} -4  \cos \phi \\ 4 \sin \phi \cot \theta \end{bmatrix}
= \alpha \begin{bmatrix} 0 \\ 2 \end{bmatrix}  + \beta
\begin{bmatrix} -2 \sin \phi \\ -2 \cos \phi \cot \theta \end{bmatrix}
\Rightarrow \alpha =  \frac{ 2 \cot \theta }{ \sin \phi }
\eeq
$\alpha=0$ implies $\theta = \pi/2$, the equator of the Bloch sphere. $\alpha$ changes sign at $\phi=0, \pm \pi$.

For the planar dynamics introduced in Section \ref{sec:OCT_planar}, we identify $\mathbf{f} = 2 \partial_\phi$ and $\mathbf{g} = -2 \sin\phi \, \partial_\theta
- 2 \cos \phi \cot\theta \,  \partial_\phi$. Using Eq.~(41) of Ref.~\cite{PhysRevA.100.022327} one gets $u^*_\text{sing}=0$. Define $\mathbb{X} = \mathbf{f} - u_\text{max} \mathbf{g}$ and $\mathbb{Y} = \mathbf{f} + u_\text{max} \mathbf{g}$, one gets $L_\mathbb{X} (\alpha) = -4  u_\text{max} < 0$ and $L_\mathbb{Y} (\alpha) = +4  u_\text{max}>0$. Following the discussion in Section IV.B in Ref.~\cite{PhysRevA.100.022327} [see also Chapter 2.9.2 in Ref.~\cite{book:GeometricOptimalControl}], $\alpha=0$ is indeed the ``fast singular arc'' that is allowed in the time-optimal protocol.

\section{Heuristics of odd harmonics} \label{app:odd_freq}

Schr\"odinger's equation for Eq.~\eqref{eqn:H_2L} 
in the rotating frame $| \psi(t) \rangle = e^{-i \omega_0 t \frac{ \sigma_z}{2}  } | \tilde{\psi}(t) \rangle$ is
\beq
i \partial_t \begin{bmatrix} \tilde{C}_1 \\ \tilde{C}_2 \end{bmatrix}
=\begin{bmatrix}0 & u(t) e^{+i \omega_0 t}  \\ u(t) e^{-i \omega_0 t} & 0 \end{bmatrix}
 \begin{bmatrix} \tilde{C}_1 \\ \tilde{C}_2 \end{bmatrix}.
\eeq
where $ [\tilde{C}_1, \tilde{C}_2]^T = | \tilde{\psi}(t) \rangle$. We consider $u(t) = V_1 \cos(\omega t) + \sum_{N=2} V_N \cos( N \omega t)$ with the initial condition $\tilde{C}_1(0)=0$,  $\tilde{C}_2(0)=1$. When $t$ is small, $
i \partial_t \tilde{C}_1 = u(t) e^{+i \omega_0 t} \tilde{C}_2
\approx \big[ V_1 \cos(\omega t) + \sum_{N=2} V_N \cos( N \omega t) \big] e^{+i \omega_0 t}$; a direct integration gives
\beq
\begin{aligned}
\tilde{C}_1(t) & \approx
\frac{V_1}{2} \bigg[ \frac{ 1 - e^{i(\omega_0 - \omega)t} }{ \omega_0 - \omega } + \frac{ 1 - e^{i(\omega_0 + \omega)t} }{ \omega_0 + \omega } \bigg] \\
& + \sum_N \frac{V_N}{2} \bigg[ \frac{ 1 - e^{i(\omega_0 - N \omega)t} }{ \omega_0 - N \omega } + \frac{ 1 - e^{i(\omega_0 + N \omega)t} }{ \omega_0 + N \omega } \bigg]
\end{aligned}
\eeq
Near resonance $\omega_0 \sim \omega$, the dominant contribution is the first term. As terms in the square bracket always come in pair, to have any effect on the second term ($\sim (\omega_0 + \omega)^{-1}$), we need to add $N=3$ terms. Similarly to have any effect on the $\sim (\omega_0 + 3 \omega)^{-1}$ term, we need to add $N=5$ terms.  For this reason the odd harmonics of the fundamental frequencies are expected to be dominant. It is reminded that the OCT analysis leading to Eq.~\eqref{eqn:optimal_bang_bang_form} consistently implies the dominant contributions are from odd harmonics of $\omega_\text{eff}$ whose value is close to $\omega_0$.

\bibliography{Quantum_control_loss}

\begin{thebibliography}{82}
\bibitem[{not({\natexlab{a}})}]{note_computation}
\bibinfo{note}{Note that the dynamics in planar form is used only for identify
  important arcs/trajectories on the Bloch sphere; it is highly non-linear and
  thus not practical for the numerical integration.}
\bibitem[{not({\natexlab{b}})}]{note_gatetime}
\bibinfo{note}{The gate time $T$ is chosen to be slightly shorter than $T^*$ so
  that $|\lambda_0|$ is small but non-zero. At $T=T^*$ both $|\lambda_0|$ and
  $\Phi$ are zero.}
\end{thebibliography}


\end{document}